\documentclass[12pt, a4paper, logo, copyright, nonumbering]{xds}

\usepackage{subcaption}
\usepackage[square,numbers,sort&compress]{natbib}
\usepackage{dblfloatfix}
\usepackage{ulem}
\usepackage{float}
\usepackage{caption}
\usepackage{dramatist}
\usepackage{xspace}
\usepackage{pifont} 
\usepackage{multirow}
\usepackage{tcolorbox}
\usepackage{xltabular}
\usepackage{longtable}
\usepackage{mdframed}
\usepackage{hyperref}
\usepackage{tocloft}
\usepackage{amsfonts}
\usepackage{amsmath}
\usepackage{amssymb}
\usepackage{lineno}
\usepackage{multirow}
\usepackage{adjustbox}

\usepackage[bottom]{footmisc}

\usepackage{CJKutf8}
\usepackage{setspace}

\usepackage{graphicx}

\usepackage{dsfont}
\usepackage{array} 
\usepackage{tabularx} 
\usepackage{xcolor} 

\usepackage{lipsum}  
\usepackage{multicol} 

\makeatletter
\def\@BTrule[#1]{%
  \ifx\longtable\undefined
    \let\@BTswitch\@BTnormal
  \else\ifx\hline\LT@hline
    \nobreak
    \let\@BTswitch\@BLTrule
  \else
     \let\@BTswitch\@BTnormal
  \fi\fi
  \global\@thisrulewidth=#1\relax
  \ifnum\@thisruleclass=\tw@\vskip\@aboverulesep\else
  \ifnum\@lastruleclass=\z@\vskip\@aboverulesep\else
  \ifnum\@lastruleclass=\@ne\vskip\doublerulesep\fi\fi\fi
  \@BTswitch}
\makeatother

\addto\extrasenglish{
}

 {\begin{list}{}%
         {\setlength{\leftmargin}{#1}}%
         \item[]%
 }
 {\end{list}}

\reportnumber{001} 
\renewcommand{\today}{}

\newcommand{\sysname}{xDeepServe}
\newcommand{\flowserve}{FlowServe\xspace}
\newcommand{\cloudmatrix}{CloudMatrix384}
\newcommand{\xccl}{XCCL}
\newcommand{\dispatch}{\texttt{dispatch}}
\newcommand{\combine}{\texttt{combine}}

\newcommand{\xds}{xDeepServe\xspace}
\newcommand{\fs}{FlowServe\xspace}

\newcommand{\ys}[1]{\textcolor{red}{Yizhou: #1}}

\newcommand{\mus}{\mbox{$\mu s$}}

\usepackage{tikz}

\title{\vspace{-0.2in}\centering Huawei Cloud Model-as-a-Service on the CloudMatrix384 SuperPod}

\author[*]{
\small xDeepServe (XDS) Team @ Huawei
}

\begin{abstract}

\textbf{Abstract.}
Scaled-out MoE LLMs and scaled-up SuperPods create new systems challenges for production Model-as-a-Service (MaaS), requiring disaggregation, low-latency communication, and decentralized serving. This report presents \sysname, the production serving system behind Huawei Cloud's MaaS offering on \cloudmatrix, a 48-server SuperPod with 384 Ascend 910C chips connected by a high-bandwidth UB fabric and global shared memory.
It serves models including DeepSeek, Kimi, GLM, Qwen, and MiniMax, among others.

\sysname\ is built around Transformerless, a disaggregated execution architecture that decomposes transformer inference into modular units---attention, feedforward, and MoE---and supports disaggregated Prefill-Decode and MoE-Attention deployments.
To enable disaggregation, we develop \xccl, a memory-semantic communication layer providing microsecond-level point-to-point and scalable all-to-all primitives, and we extend \flowserve\ with decentralized DP groups and techniques to mitigate stragglers and synchronization variance.
In a peak decoding configuration, \sysname\ reaches 2400 tokens/s per Ascend 910C chip at $\sim$50\,ms time-per-output-token (TPOT).

\vspace{0.3in}

\end{abstract}

\begin{document}
\begin{CJK*}{UTF8}{gbsn}

\maketitle

\begin{figure}[H]
  \centering
  \includegraphics[width=0.95\textwidth]{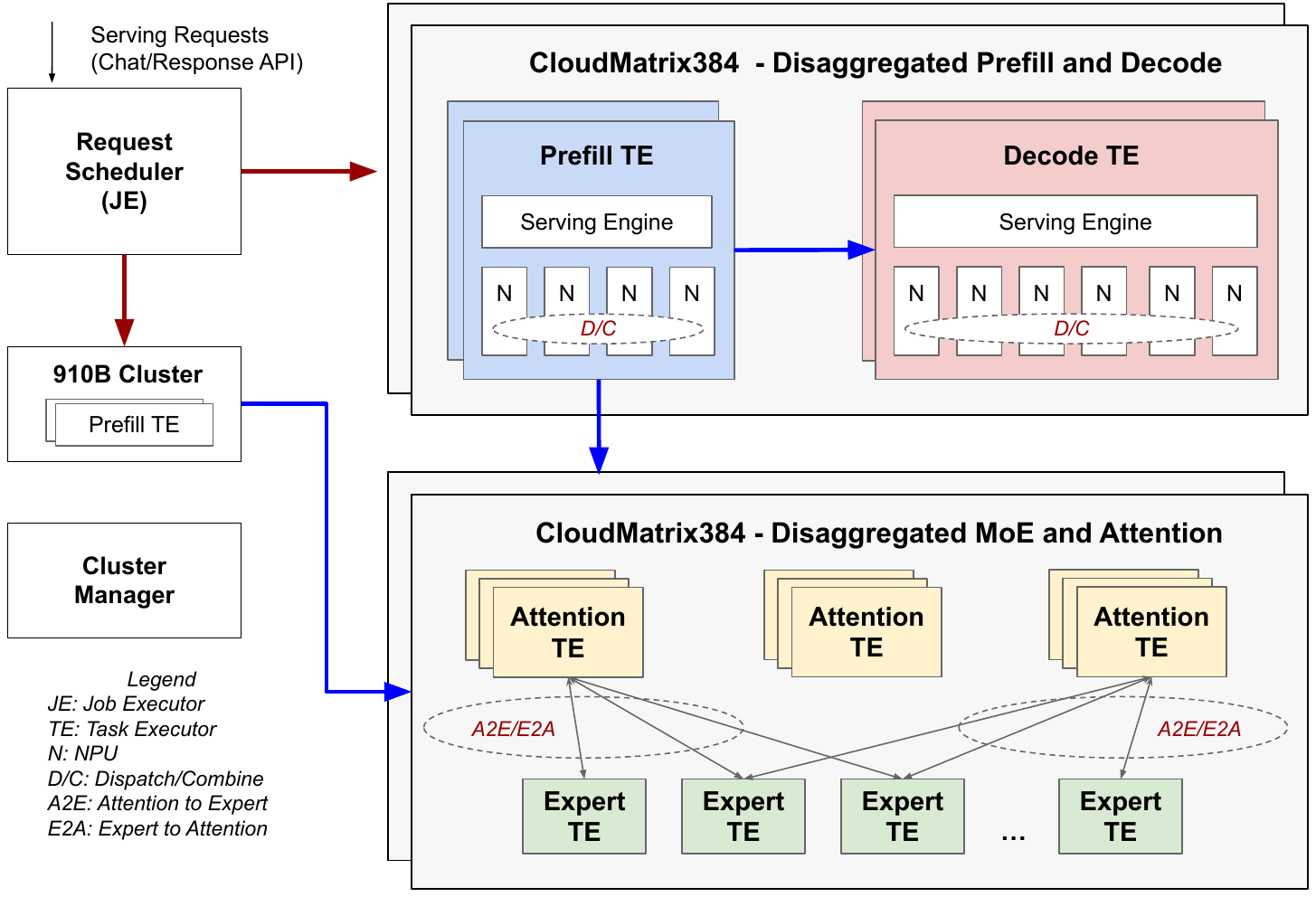}
  \caption{\centering\textbf{\xds Architecture over \cloudmatrix\ SuperPod.}}
  \label{fig-system-overview}
\end{figure}

\newpage
\tableofcontents
\newpage

\section{Introduction}

Large language models (LLMs) keep growing in size and complexity.
To improve quality and efficiency, many recent models, such as DeepSeek~\cite{liu2024deepseek-v3}, Kimi K2~\cite{kimi-k2}, and Qwen~\cite{yang2025qwen3technicalreport}, adopt the Mixture-of-Experts (MoE) architecture. MoE scales out model capacity by activating only a small subset of experts per token, reducing compute cost while keeping high model quality. At the same time, AI hardware is scaling up. Huawei’s \cloudmatrix\ connects 384 Ascend 910C chips with hundreds of GB/s of high-speed interconnect bandwidth, offering a tightly coupled memory and compute environment across the entire pod.

When these two trends meet---scaled-out MoE models and scaled-up SuperPods---new challenges emerge.
MoE models require fine-grained expert routing, synchronization, and load balancing across hundreds of NPU devices.
SuperPod hardware enables global shared memory and uniform low-latency access across hundreds of NPUs, but traditional LLM serving systems are not built to exploit these properties.
To unlock the full potential of SuperPod-scale infrastructure, we need a new system that runs disaggregated workloads, maintains low tail latency, and scales efficiently across hundreds of NPUs.

This paper presents \textbf{\sysname}, Huawei Cloud’s in-house LLM serving system for SuperPod-scale deployment.

We advocate resource disaggregation~\cite{legoos-osdi18} as a foundational design principle for large-scale serving. Disaggregation decouples system components into independently scalable units, improves fault isolation, and enables flexible system evolution. To realize these benefits, we introduce \textit{Transformerless}, an architecture that breaks transformer-based LLMs into modular components---attention, feedforward, and MoE---and runs each module independently on dedicated NPUs. This design decouples compute and memory, reduces interference across stages, and supports separate scaling.

We implement two forms of disaggregation.

First, to reduce interference between compute-bound prefill and memory-bound decode phases, we use a disaggregated Prefill-Decode architecture (\S\ref{sec:disagg-pd}).
While prior systems like Splitwise~\cite{patel2023splitwise}, TetriServe~\cite{tetriserve-arxiv24}, and DistServe~\cite{zhong2024distserve} explored this direction, our design targets the unique demands of large MoE models with expert parallelism on SuperPod-scale hardware. This disaggregated pipeline is further enhanced by heterogeneous deployment: we run prefill on both Ascend 910B and 910C NPUs, while decode runs exclusively on CloudMatrix’s 910C nodes to leverage high-bandwidth interconnects. This design balances cost and performance, while achieving high throughput and low latency at scale.

Second, we further disaggregate MoE-Attention by running MoE experts and attention on separate NPUs to reduce resource contention and improve utilization (\S\ref{sec:disaggg-ma}). While prior work explored this idea on smaller models, we demonstrate this design at SuperPod scale by deploying DeepSeek-V3/R1 across 768 Ascend 910C dies, with 288 handling MoE and 480 handling attention.
This separation introduces a compute imbalance: MoE is stateless and scales with batch size, while attention maintains KV cache and scales with sequence length. To address this, we introduce three techniques: (1) a trampoline-based \texttt{A2E}/\texttt{E2A} routing protocol to handle asymmetric NPU allocation, (2) a new \textit{DP domain} abstraction to enable inter-group scheduling without shrinking batch sizes, and (3) persistent kernel scheduling across concurrent compute and communication streams to eliminate CPU overhead. These designs allow us to fully utilize both MoE and attention NPUs while maintaining sub-200\,\mus dispatch latency across the entire SuperPod.

To support fully disaggregated execution, we propose \xccl, a novel communication library that provides memory-semantic point-to-point and all-to-all primitives over \cloudmatrix’s shared-memory fabric.
\xccl\ supports microsecond-level latency and high concurrency across more than 300,000 NPU pairs.
We use it to build KV cache transfer, MoE dispatch/combine, and MoE-Attention (\S\ref{sec:communication}).

\sysname\ further builds a scalable serving engine at SuperPod scale, called \flowserve. It introduces the \textit{Data Parallel (DP) group} abstraction, where each group manages its own execution pipeline---including tokenization, SPMD execution, caching, and networking---without relying on central coordination. 
\flowserve\ scales to hundreds of NPUs by eliminating single points of scalability bottlenecks and applying system-level techniques to reduce latency under MoE barriers such as expert \dispatch\ and \combine.
These include load-aware request routing among DPs, proactive garbage collection to reduce jitter, and MoE expert load balancing.
To improve efficiency and hardware utilization, \flowserve\ supports Multi-Token Prediction (MTP) and INT8 quantization, enabling high-throughput inference under tight SLA constraints.

To ensure reliability at SuperPod scale, \sysname\ incorporates reliability mechanisms across failure detection and recovery paths.
It uses a multi-level heartbeat and link-probing system to detect faults ranging from hung processes to silent KV-transfer stalls.
For recovery, \sysname\ evolves from coarse-grained cluster restarts to fine-grained component-level resilience. It supports independent failover of prefill and decode stages, dynamic reconfiguration of expert ranks, and selective token recomputation on transient network or memory faults. These designs allow the system to maintain availability and throughput even under hardware disruptions, a critical capability for large-scale MoE serving.

\sysname\ is deployed in production to serve large-scale DeepSeek~\cite{liu2024deepseek-v3}, GLM~\cite{glm5team2026glm5vibecodingagentic}, Kimi~\cite{kimi-k2}, and Qwen~\cite{yang2025qwen3technicalreport} models. In our peak decoding configuration, it achieves 2400 tokens/s per Ascend 910C chip at $\sim$50\,ms time-per-output-token (TPOT). In production, we target a TTFT SLA under 2\,s and a TPOT SLA of 35\,ms in most cases.

\noindent\textbf{Contributions.} We make the following contributions:
\begin{itemize}
    \item We present \sysname, a production MaaS serving system for running large MoE LLMs on \cloudmatrix\ SuperPods.
    \item We describe Transformerless, including disaggregated Prefill-Decode and disaggregated MoE-Attention deployments at SuperPod scale.
    \item We introduce \xccl, a communication layer over \cloudmatrix’s distributed shared memory that provides microsecond-level point-to-point and all-to-all primitives for serving.
    \item We present the SuperPod-scale serving design of \flowserve, including decentralized DP groups, load-aware scheduling, expert load balancing, and reliability mechanisms.
\end{itemize}

\noindent
In the rest of the paper, we first describe the architecture of \cloudmatrix\ in \S\ref{sec:bg-cm384}, followed by the low-level communication library in \S\ref{sec:communication}. We then present the serving engine in \S\ref{sec:flowserve} and the Transformerless serving architecture in \S\ref{sec:transformerless}. Finally, we discuss our reliability mechanisms in \S\ref{sec:reliability}.

\section{Background}

\subsection{\sysname}

\sysname\ is Huawei Cloud's fully-hosted, serverless LLM serving platform, designed to support large-scale, multi-tenant workloads. It evolves from the DeepServe system introduced in our previous work~\cite{hu2025deepflow}, with new capabilities targeting LLM serving at the scale of \cloudmatrix.
\sysname\ has been in production for over a year, operating on a large Ascend NPU cluster. It provides industry-standard APIs for fine-tuning, agent hosting, and model serving. Motivated by the growing demand for generative AI services (e.g., ChatGPT and DeepSeek), \sysname\ addresses key production challenges, including heterogeneous workload durations, distributed stateful execution, and dynamic resource demands.

To tackle these challenges, \sysname\ introduces four core components~\cite{hu2025deepflow}.  
First, it adopts a serverless abstraction based on a request-job-task model to manage diverse AI workloads, spanning post-training and inference tasks.  
Second, it integrates a custom serving engine, \flowserve, which follows a microkernel-inspired architecture, supports NPU-centric execution, and uses SPMD-style parallelism for high-performance LLM serving.  
Third, it includes scheduling policies designed to support both Prefill-Decode (PD) disaggregated and PD-colocated deployment modes.  
Fourth, it incorporates system-level optimizations, such as pre-warmed pods, DRAM preloading, and NPU fork, to enable rapid elasticity, scaling to 64 instances within seconds.

While our previous paper~\cite{hu2025deepflow} introduced the core architecture of \sysname, this report focuses on the new challenges and techniques arising from deployment on the \cloudmatrix\ SuperPod. We present our latest system designs and methods for leveraging its scaled-up capabilities to enable next-generation LLM serving.

\subsection{CloudMatrix384}
\label{sec:bg-cm384}

%
%

\begin{figure}[t]
    \centering
    \includegraphics[width=\textwidth]{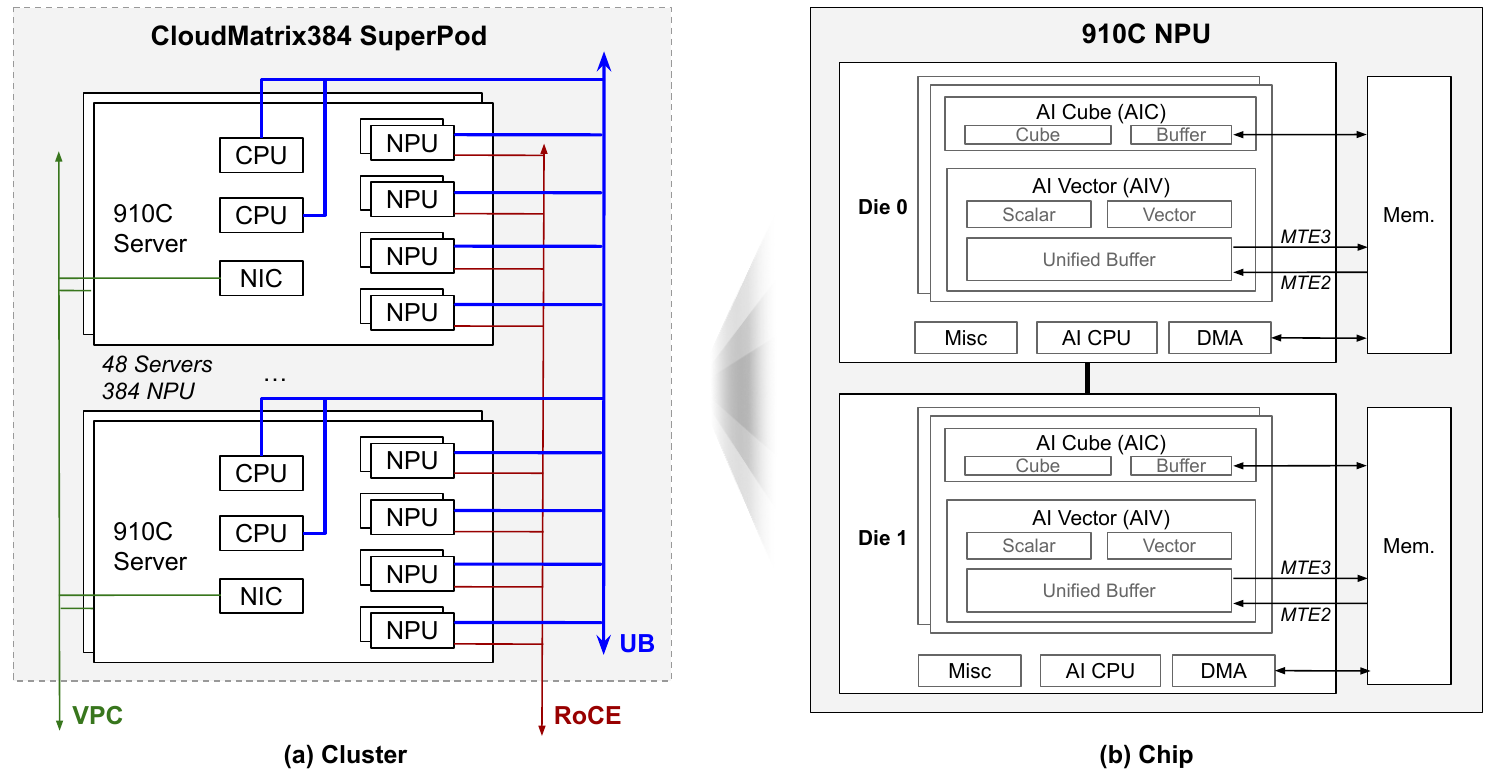}
    \caption{\textbf{An Overview of CloudMatrix384 SuperPod and Ascend 910C NPU Chip.}
    \textit{A \cloudmatrix\ SuperPod has 48 servers and 384 Ascend 910C NPU chips in total.
    A single 910C NPU chip has two dies interconnected via high-bandwidth NoC.
    The chip uses the decoupled DaVinci architecture, including independent AI cube, AI vector, etc.}
    }
    \label{fig-cloudmatrix}
\end{figure}

Figure~\ref{fig-cloudmatrix} shows an overview of \cloudmatrix\ and the 910C NPU chip.

A \cloudmatrix\ SuperPod is a single scale-up domain composed of 48 servers, delivering hundreds of PFLOPs of FP16 compute, several terabytes of on-chip memory, and terabytes per second of memory bandwidth.
Each server in the SuperPod is equipped with multiple CPUs, multiple NICs, and 8 Ascend 910C NPU chips.
Crucially, the SuperPod integrates three types of network fabrics. The first is a traditional VPC network that connects servers to external systems or cloud services.
The second is a scaled-out RoCE network that interconnects all NPU chips and can be extended across multiple SuperPods and 910B servers.
The third is a scaled-up UB network that makes \cloudmatrix\ a SuperPod.
It connects all CPUs and NPUs within the SuperPod in an all-to-all fashion, offering several times higher bandwidth than RoCE.

The scaled-up UB network enables several novel properties.
First, it provides a global shared memory address space spanning all CPU DRAM and NPU on-chip memory, allowing any chip to access the memory of any other chip.
This access supports two semantics: traditional DMA and memory semantics, which will be detailed later.
Second, the system eliminates intra-server NUMA locality---CPUs have uniform access latency to all eight NPUs within a single server.

An Ascend 910C NPU chip consists of two dies interconnected via an on-chip network.
%
%
Each die follows the DaVinci architecture~\cite{ascend-npu-hotchips19,ascend-c-programming-manual,cloudmatrix384}, comprising an equal number of AI Cube (AIC) and AI Vector (AIV) cores, several AI CPUs, and multiple DMA engines.
AIC cores execute matrix-oriented kernels, while AIV cores handle vector operations. Both core types include scalar units and support independent control flow, with data exchange via external on-chip memory.
Each AIV core includes a KB-level unified buffer and two Memory Transfer Engines (MTEs), codenamed MTE2 and MTE3, which transfer data between on-chip memory and the unified buffer: MTE2 loads data from on-chip memory into the buffer, and MTE3 writes data back to on-chip memory.

The MTE2/MTE3 engines in AIV cores, along with the DMA engines, are fundamental to enabling global shared memory in \cloudmatrix.
Both MTE2/MTE3 and DMA engines on an NPU can read from or write to the on-chip memory of any other NPU in the SuperPod via the UB fabric. This capability underpins our customized communication library for serving (\S\ref{sec:communication}).
In general, MTE2/MTE3 are used for low-latency communication, while DMA engines are preferred for high-throughput transfers.
This is because MTE2/MTE3 provide a memory-semantic API and move data between the AIV's unified buffer and on-chip memory, with transfers limited to the buffer size of hundreds of KBs. In contrast, DMA engines use DMA semantics and support bulk data movement up to several GBs.
For more details, refer to the Ascend C Programming Manual.

\subsection{NPU Execution Mode}

%
%
%

\begin{figure}[t]
    \centering
    \includegraphics[width=\textwidth]{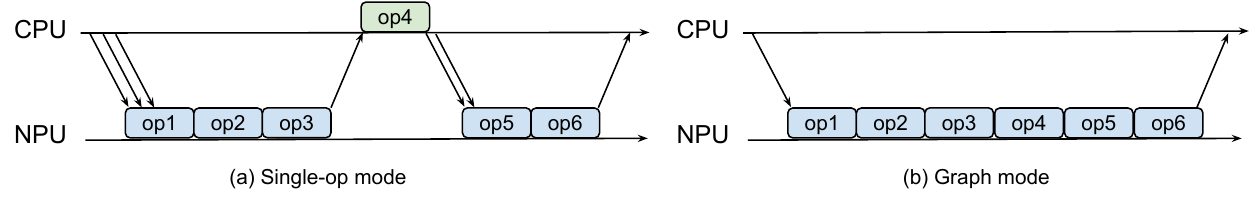}
    \caption{\textbf{Compare Single-Op and Graph Mode Execution on Ascend NPU.}}
    \label{fig-graph-mode}
\end{figure}

We describe the two execution modes supported by Ascend NPUs and how we apply them at different stages of LLM serving to balance flexibility and performance.
Frameworks, such as PyTorch, represent neural networks as computation graphs, where operators are nodes and data dependencies are edges. These graphs can run in two modes: \textit{single-op} and \textit{graph}.

\begin{itemize}
\item \textbf{Single-op Mode.}  
This is PyTorch’s default execution mode. Each operator is dispatched to the NPU operator queue
upon Python invocation, and its result is accessible on the next synchronization. This dynamic style supports arbitrary control flow, dynamic tensor shapes, and native debugging, and it allows limited CPU-NPU concurrency when dependencies permit. However, it introduces overhead because each operator launch incurs host-device communication, and the NPU may stay idle if the operator execution time
is lower than the dispatch time of the next operator. That said, for computation-heavy operators, this cost is acceptable. Therefore, we use this mode during prefill to handle dynamic input shapes.

\item \textbf{Graph Mode.}  
Graph mode traces and compiles a static computation graph, which is dispatched as a single unit to the NPU. This removes per-operator launch overhead and maximizes device utilization. Huawei’s TorchAir (built on PTA) supports graph-mode inference on Ascend NPUs. As shown in Figure~\ref{fig-graph-mode}(b), the CPU can dispatch the entire graph to the NPU in one kernel launch. We use graph mode during decode for optimal performance, as input shapes are small and mostly static.
\end{itemize}

\section{Communication Library over CloudMatrix's Distributed Shared Memory}
\label{sec:communication}

A high-performance communication library is essential to fully harness the scaled-up UB fabric in \cloudmatrix.
We introduce \xccl, a communication library purpose-built for LLM serving on SuperPod.
\xccl\ proposes distributed memory protocols over \cloudmatrix's global shared memory to construct efficient networking primitives.
Its protocol design resembles classical far-memory systems based on one-sided RDMA verbs~\cite{farm-nsdi14, clover-atc20}.

We describe the point-to-point APIs used in disaggregated Prefill-Decode (\S\ref{sec:xccl-p2p}), the dispatch and combine APIs for expert parallelism (EP) (\S\ref{sec:xccl-dispatch}), and the A2E/E2A APIs for disaggregated MoE-Attention (\S\ref{sec:xccl-a2e}).

\subsection{Point-to-Point Primitives}
\label{sec:xccl-p2p}

%
%

\begin{figure}[t]
	    \centering
	    \includegraphics[width=\textwidth]{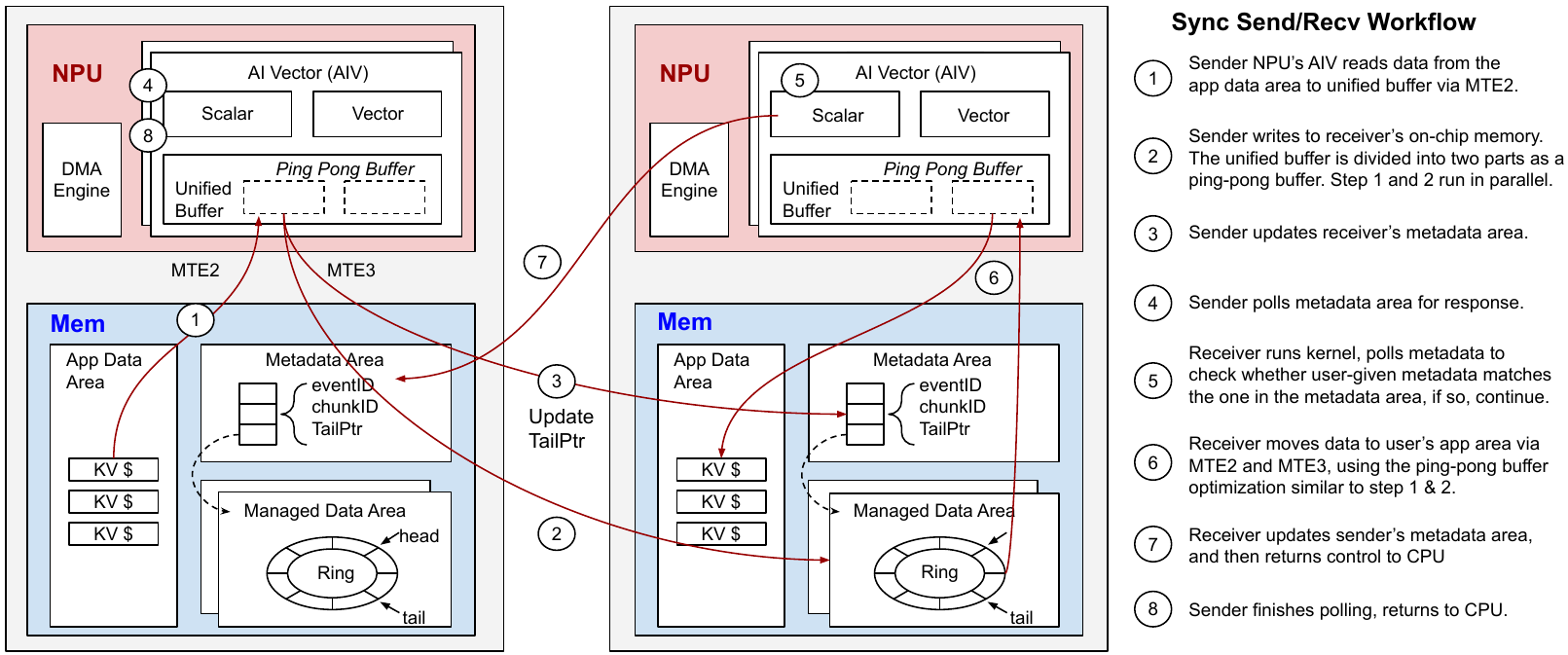}
	    \caption{\textbf{Distributed Send/Receive Workflow.} \textit{We show two NPUs and the distributed memory transfer protocol. We only show memory-semantic-based transfer using MTE2/MTE3 while remote memory copies can also be performed using the DMA engine. We also have a zero-copy version in which the send and receive kernels directly manipulate the app data area.}}
	    \label{fig-xccl-send}
\end{figure}

This section describes the design and implementation of point-to-point APIs such as  \texttt{send} and \texttt{receive} on the \cloudmatrix\ SuperPod.
These APIs are essential for serving, enabling efficient data transfer in scenarios such as disaggregated prefill and decode, sequence parallelism, and fast scaling via \texttt{npu-fork}~\cite{hu2025deepflow}.
For simplicity, we use the transfer of KV cache~\cite{hu2025epic} from prefill NPU to decode NPU to illustrate the \texttt{send}/\texttt{receive} design. 
\xccl's \texttt{send} and \texttt{receive} operations can occur between any pair of NPUs in \cloudmatrix, hence our design can scale to roughly 300K potential pairs.

We now describe the design details.

\textbf{Data structure.}
Each NPU’s on-chip memory is partitioned into three areas: the app data area, the metadata area, and the managed data area.
The \textbf{app data area} is reserved for application-specific data such as KV cache, used directly by our serving engine.
The \textbf{metadata area} stores control information required for sanity checks and to implement the distributed memory protocol.
It consists of metadata fields, one for each pair of AIV cores to enable parallelism among AIV cores.
Given that \cloudmatrix\ includes 384 NPU chips, each with 2 dies and up to 48 AIV cores per die, the total number of fields is $384 \times 2 \times 48 \times 2 \approx 74\text{K}$. Each 32-byte field includes a user-defined \texttt{eventID} for sanity checking, a kernel-generated \texttt{chunkID} for tracking chunked transfers, and a \texttt{tailPtr} pointing to the ring buffer in the data area. The total metadata size is set to 4\,MB.
The \textbf{managed data area} is used for data exchange between NPUs. A dedicated ring buffer is maintained for each NPU pair, with a fixed number of slots of fixed size per buffer.
Both the metadata and managed data areas are managed by \xccl. The current implementation is not zero-copy; data is copied between the app data area and the managed data area when \xccl\ APIs are invoked.

\textbf{Distributed Memory Protocol.}
We design a distributed protocol to transfer data from sender to receiver over \cloudmatrix's global shared memory, illustrated in Figure~\ref{fig-xccl-send}.
\textbf{Step 1:} The sender’s serving engine invokes \xccl’s \texttt{send}, passing the source buffer in the app data area (e.g., KV cache), an \texttt{eventID} (e.g., number of sends), the receiver NPU’s ID, and the number of AIV cores to use. \xccl\ launches a kernel on the sender NPU.
The send kernel uses MTE2 to copy data from the app data area to each AIV’s unified buffer in parallel.
\textbf{Step 2:} The send kernel then reads the destination's on-chip memory address from its metadata area and uses MTE3 to transfer data from the unified buffer to the receiver's managed data area. Each unified buffer operates in a ping-pong fashion, allowing MTE2 and MTE3 to run concurrently.
Alternatively, the send kernel can copy data from the app data area to the receiver's managed data area using the DMA engine, which bypasses the unified buffer but incurs higher startup latency.
\textbf{Step 3:} The send kernel updates the receiver’s \texttt{tailPtr} in the metadata area via MTE3, indicating the amount of data transferred.
\textbf{Step 4:} The send kernel then busy-polls its local metadata area for an acknowledgment from the receiver.
\textbf{Step 5:} Meanwhile, the receiver’s serving engine invokes \xccl’s \texttt{receive}, passing the destination buffer in the app data area, the \texttt{eventID}, the sender NPU’s ID, and the number of AIV cores. \xccl\ launches a kernel on the receiver NPU, which polls the metadata area for new data.
\textbf{Step 6:} Upon detecting data, the receive kernel copies it from the managed data area to the app data area using MTE2 and MTE3, again in a ping-pong fashion.
\textbf{Step 7:} After the transfer completes, the receive kernel updates the sender’s metadata area to acknowledge completion, then returns control to the CPU and the \texttt{receive} caller.
\textbf{Step 8:} The send kernel detects the acknowledgment, returns control to the CPU, and completes the \texttt{send} call.
In addition to this synchronous protocol, \xccl\ supports an asynchronous mode in which send and receive kernels avoid busy polling.

\begin{figure}[t]
    \centering
	    \begin{minipage}[h]{0.48\textwidth}
	        \centering
	        \includegraphics[width=\textwidth]{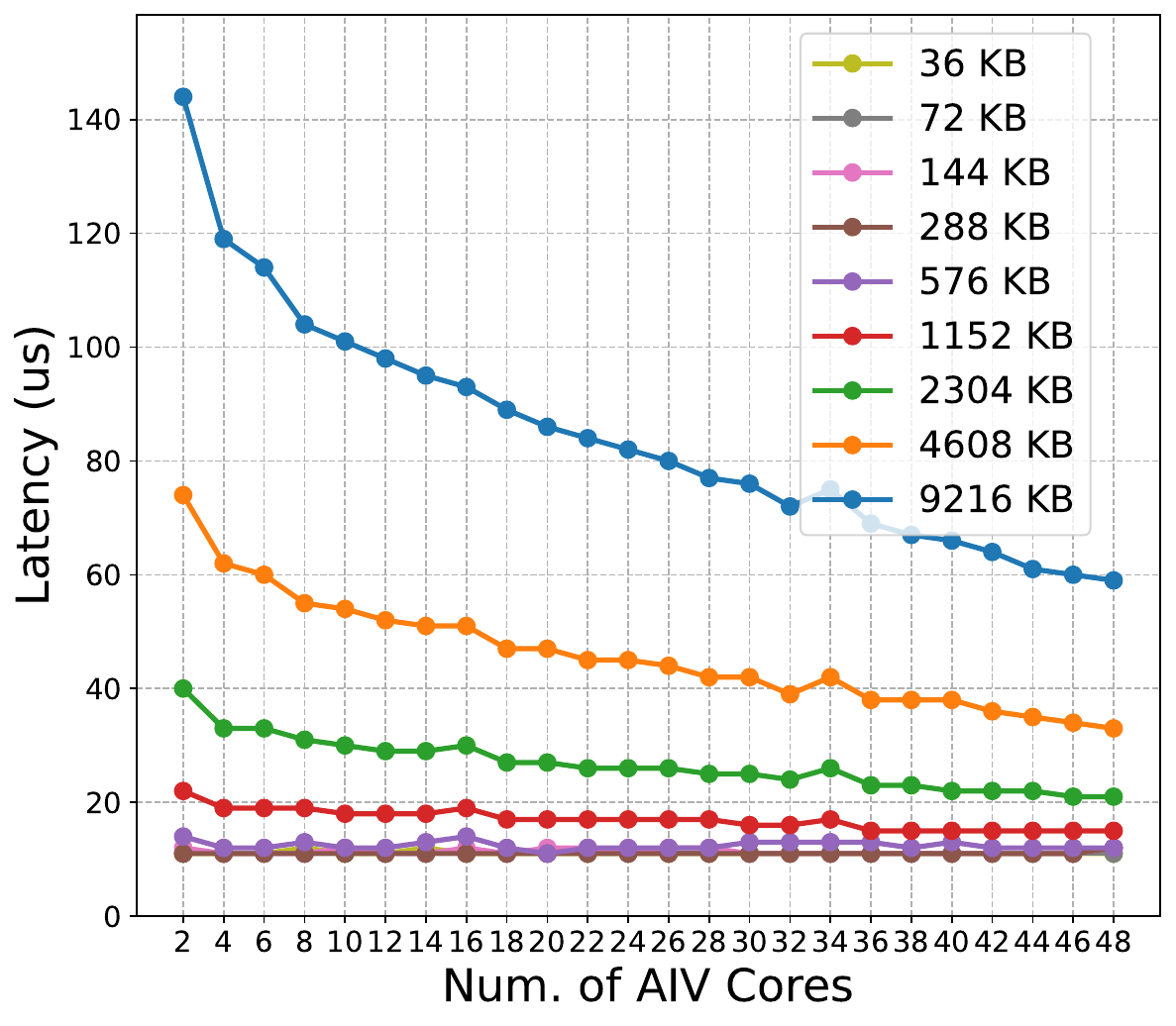}
	        \caption{\textbf{Evaluation of Send/Receive.} \textit{We vary the data size and the number of AIV cores used for a single send/receive pair.}}
	        \label{fig-xccl-send-eval}
	    \end{minipage}
    \hfill
	    \begin{minipage}[h]{0.48\textwidth}
	        \centering
	        \includegraphics[width=\textwidth]{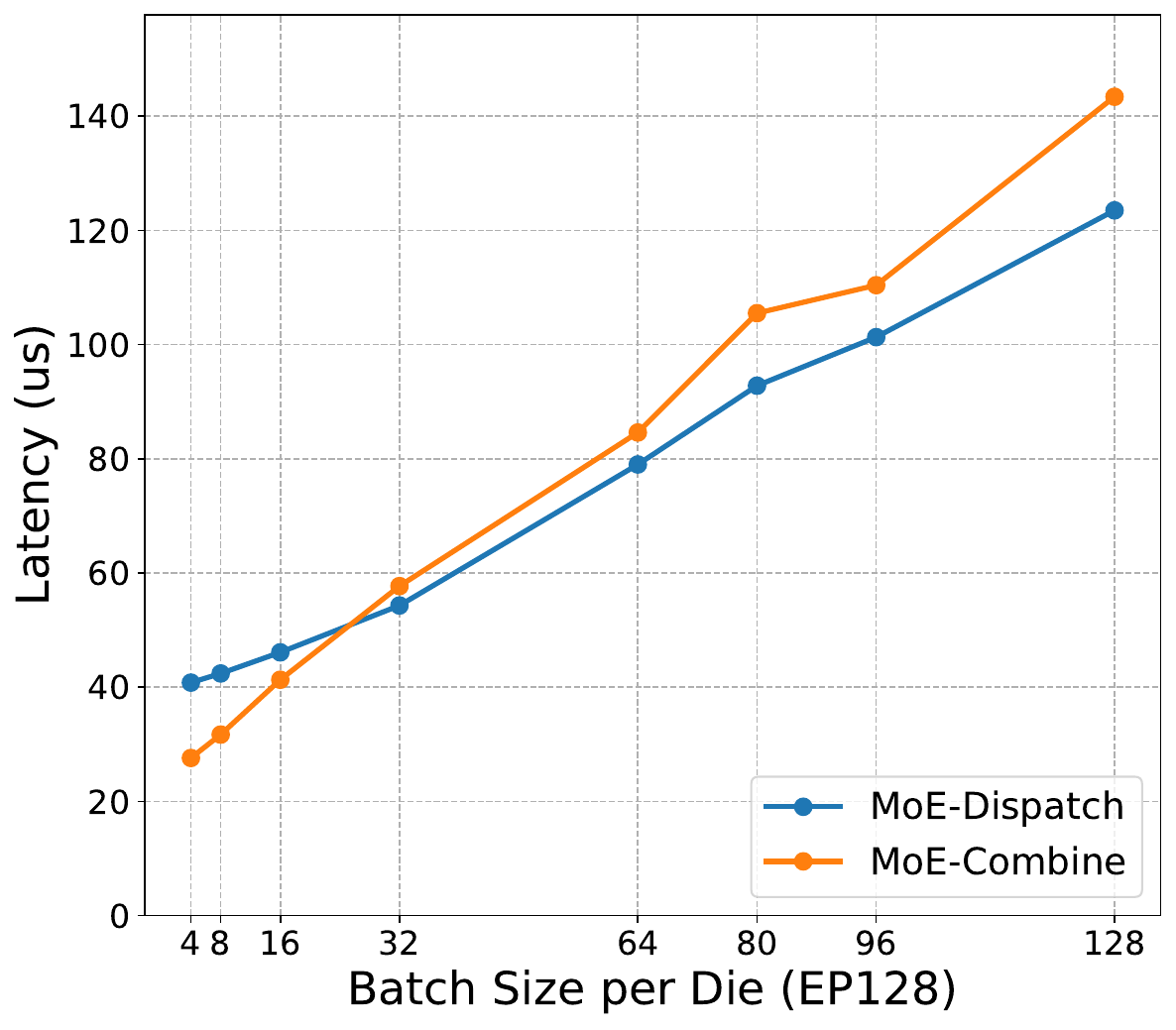} 
	        \caption{\textbf{Evaluation of Dispatch/Combine.} \textit{We vary the batch size per die with a fixed EP128. Dispatch includes a quantization step.}} 
	        \label{fig-xccl-dispatch-combine-eval}
	    \end{minipage}
\end{figure}

\textbf{Performance.}
Our \texttt{send} and \texttt{receive} primitives achieve microsecond-level latency.
Figure~\ref{fig-xccl-send-eval} evaluates their performance under varying data sizes and numbers of AIV cores per kernel invocation.  
We measure end-to-end latency, from the start to the completion of \texttt{send}, including all protocol steps described earlier.  
We randomly select two NPU dies at different SuperPod servers, leveraging the uniform bandwidth of the scaled-up UB fabric.
As shown in Figure~\ref{fig-xccl-send-eval}, for payloads smaller than 1\,MB, latency remains under 20\,\mus even with just 2 AIV cores. As the data size increases, performance scales with parallelism: transferring 9\,MB using all 48 AIV cores is more than 2.5$\times$ faster than using only 2.

\subsection{All-to-All Communication for Colocated MoE-Attention}
\label{sec:xccl-dispatch}
\begin{figure}[h]
    \centering
    \includegraphics[width=\textwidth]{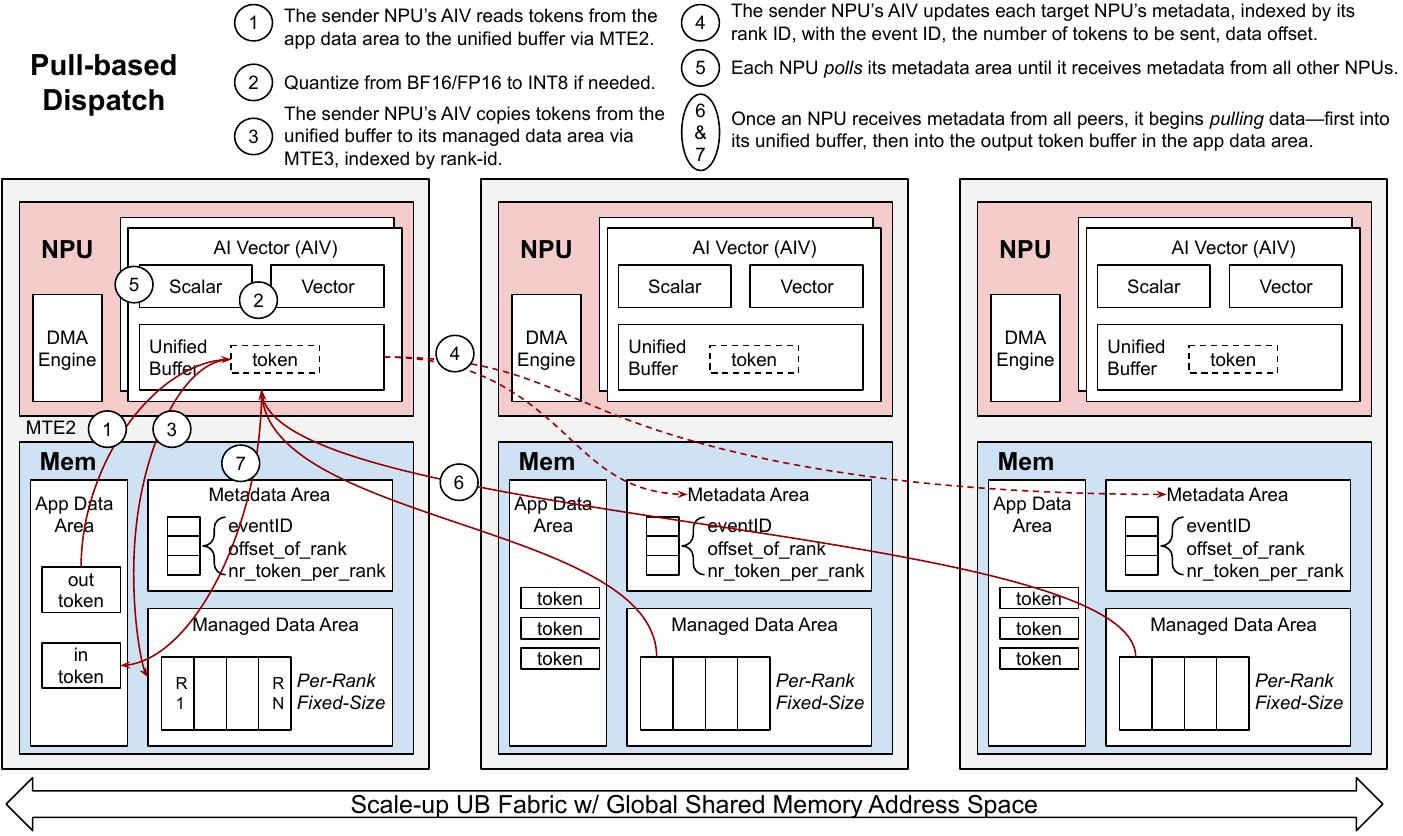}
   \caption{\textbf{Pull-based Dispatch based on UB's Global Shared Memory.} \textit{Dashed red lines are metadata transfer. Bulk data transfer in step 6 and 7 can also use the DMA engine. A single dispatch kernel can use multiple AIV cores. All NPUs involved in a dispatch run the same seven steps described above. We show three NPUs for simplicity.}}
    \label{fig-xccl-dispatch}
\end{figure}

This section presents the design and implementation of two all-to-all communication primitives---\texttt{dispatch} and \texttt{combine}---used in large-scale expert parallelism (EP).
In EP, \texttt{dispatch} routes each token’s hidden state to its top-$k$ experts based on gating scores, while \texttt{combine} aggregates expert outputs, weighted by the same scores, into a unified tensor. Together, we find these primitives account for at least 25\% of MoE execution time.
Similar to point-to-point primitives, we implement these all-to-all operations using far memory semantics rather than traditional network-level verbs like DeepEP~\cite{deepep}.
For brevity, we detail the \dispatch\ protocol below; \combine\ follows a similar pattern.

\textbf{Data Structure.}
Similar to the point-to-point primitives, we partition each NPU’s on-chip memory into three regions.
The metadata area contains 32-byte fields, one per rank. Each field includes an event ID for sanity checking, a pointer to the corresponding rank’s buffer offset in the managed data area, and a token count received from that rank.
The max number of metadata fields is similar to that of point-to-point primitives.
The managed data area is partitioned by rank ID, with each NPU assigned a fixed-size block determined by the maximum supported batch size.

\textbf{Distributed Memory Protocol of \dispatch.}
The \dispatch\ protocol consists of two main phases: broadcasting the number of tokens each rank should receive, followed by the actual data transfer. Figure~\ref{fig-xccl-dispatch} illustrates the complete workflow.
\textbf{Step 1:} The sender's serving engine invokes \dispatch, passing the source buffer in the app data area, an event ID, and additional parameters. A kernel is launched on the sender NPU, which uses MTE2 to copy data from the app data area to each AIV's unified buffer in parallel.
\textbf{Step 2:} If quantization is enabled, the kernel converts the data from FP16/BF16 to INT8 using vector instructions.
\textbf{Step 3:} The kernel writes token data into the managed data area, partitioned by destination rank ID.
\textbf{Step 4:} The kernel updates each destination rank's metadata with the number of tokens it will receive.
\textbf{Step 5:} Each NPU polls its local metadata area until metadata from all ranks is received.
\textbf{Steps 6--7:} Once all metadata is received, the kernel pulls token data from peer NPUs into the unified buffer using MTE2, guided by received offsets and token counts. It then copies the data into the destination buffer in the app data area via MTE3.

\textbf{Performance.}
Figure~\ref{fig-xccl-dispatch-combine-eval} evaluates \dispatch\ and \combine's performance under varying batch sizes per die, using DeepSeek-R1 models with a fixed EP128 configuration.
At small batch sizes, \dispatch\ exhibits slightly higher latency than \combine\ due to the additional overhead from quantization. However, since quantization reduces data size by half, \dispatch\ becomes faster than \combine\ when the batch size per die exceeds 32.
For reference, with a batch size per die of 96 and EP128, the total global batch size across the system reaches \(96 \times 128 = 12{,}288\).

\subsection{All-to-All Communication for Disaggregated MoE-Attention}
\label{sec:xccl-a2e}

%
%

\begin{figure}[t]
    \centering
    \includegraphics[width=0.98\textwidth]{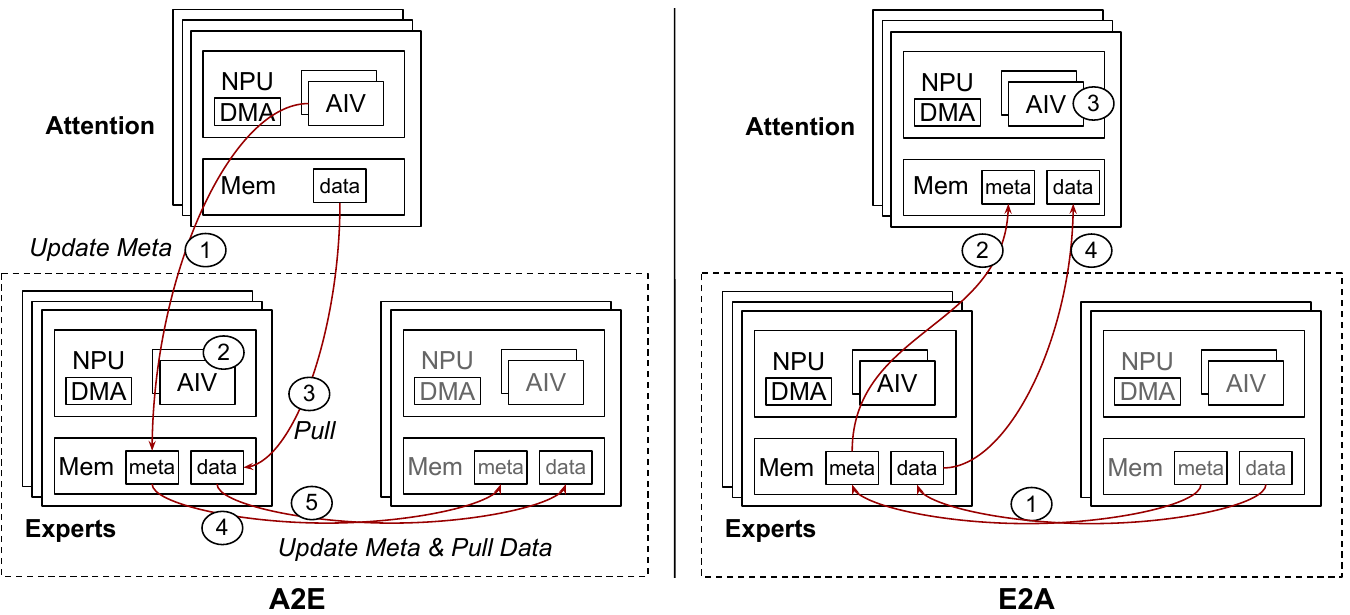}
    \caption{\textbf{Attention 2 Expert and Expert 2 Attention Primitives.} \textit{In \texttt{A2E}, the leftmost set of expert NPUs serves as trampolines, receiving data directly from the attention NPUs.  
In \texttt{E2A}, expert outputs are first routed to the trampoline NPUs, which then forward the data to the attention NPUs.  
This two-stage routing reduces metadata updates and balances communication across asymmetric NPU allocations.}}
    \label{fig-xccl-a2e-e2a}
\end{figure}

This section presents the design of two all-to-all communication primitives---\texttt{A2E} (\texttt{attention2\allowbreak expert}) and \texttt{E2A} (\texttt{expert2\allowbreak attention})---used in disaggregated MoE-Attention deployments (see \S\ref{sec:disaggg-ma}). These primitives enable efficient token routing and aggregation between attention NPUs and expert NPUs at scale.

Similar to \texttt{dispatch}, the \texttt{A2E} primitive routes each token’s hidden state from the attention NPUs to its top-$k$ expert NPUs, as determined by the gating scores. Conversely, \texttt{E2A} functions like \texttt{combine}, collecting outputs from the selected experts and merging them---weighted by the same gating scores---back at the attention NPUs. Both primitives use the same distributed shared memory infrastructure and share similar data structures with \texttt{dispatch}/\texttt{combine}.

A key difference, however, is that attention and expert modules reside on separate NPU dies, introducing asymmetry in resource allocation. In large-scale deployments, this asymmetry presents new challenges. For example, in a typical DeepSeek-R1/V3~\cite{liu2024deepseek-v3} configuration with 288 experts, we often provision 288 expert NPUs but only 160 attention NPUs (see \S\ref{sec:disaggg-ma}). In a naive pull-based \texttt{dispatch} design, each attention NPU would push metadata for all expert NPUs and wait for them to pull data. This approach quickly becomes inefficient due to the high fan-out and limited scalar throughput of each AIV core.

\textbf{Trampoline Forward}.
To address this, we introduce a novel design called trampoline forward. In this scheme, a subset of expert NPUs---equal in number to the attention NPUs---is designated as trampolines. These trampoline NPUs first receive all data from the attention NPUs and then forward it to the remaining expert NPUs. This two-stage routing reduces metadata overhead and balances traffic across the asymmetrically allocated NPUs. The complete data flow is illustrated in Figure~\ref{fig-xccl-a2e-e2a}.

\textbf{Trade-off between MTE and DMA.}
To improve communication efficiency, we employ \texttt{NPU-Direct Unified Remote Memory Access (URMA)}, a technique on Ascend NPUs similar to IBGDA on GPUs~\cite{liu2024deepseek-v3}.
\texttt{NPU-Direct URMA} enables AIV cores to issue remote memory access requests directly to the DMA engine, bypassing both the host CPU and AI CPU as shown in \S\ref{sec:bg-cm384}.
Although \texttt{NPU-Direct URMA} incurs higher startup latency compared to MTE2/MTE3, it offers three key advantages.
First, it reduces AIV consumption. By offloading memory access to the DMA engine using asynchronous semantics, more AIV resources become available for communication or computation.
Second, it is better suited for high-throughput scenarios.
MTE2/MTE3 transfers are limited by the AIV's unified buffer size, whereas the DMA engine supports transfers up to several GBs.
Third, it avoids contention with compute streams. Since MTE2 is also used for computation, using DMA helps prevent interference when compute and communication streams share the same NPU die, as discussed in \S\ref{sec:disaggg-ma}.

\textbf{Performance.}  
We evaluate \texttt{A2E} and \texttt{E2A} at SuperPod scale using a single deployment with three DP domains, each containing 160 DP groups (TP = 1), and 288 expert NPUs. We run DeepSeek-R1 with a per-die batch size of 96, resulting in a global batch size of \(96 \times 3 \times 160 = 46{,}080\). Under this configuration, \texttt{A2E} achieves a latency of 172\,\mus, while \texttt{E2A} completes in 193\,\mus. These results demonstrate that both primitives maintain low-latency execution even under high concurrency and large-scale deployment.

\section{Scalable Serving System at SuperPod-scale}
\label{sec:flowserve}

\begin{figure}[t]
    \centering
    \includegraphics[width=\textwidth]{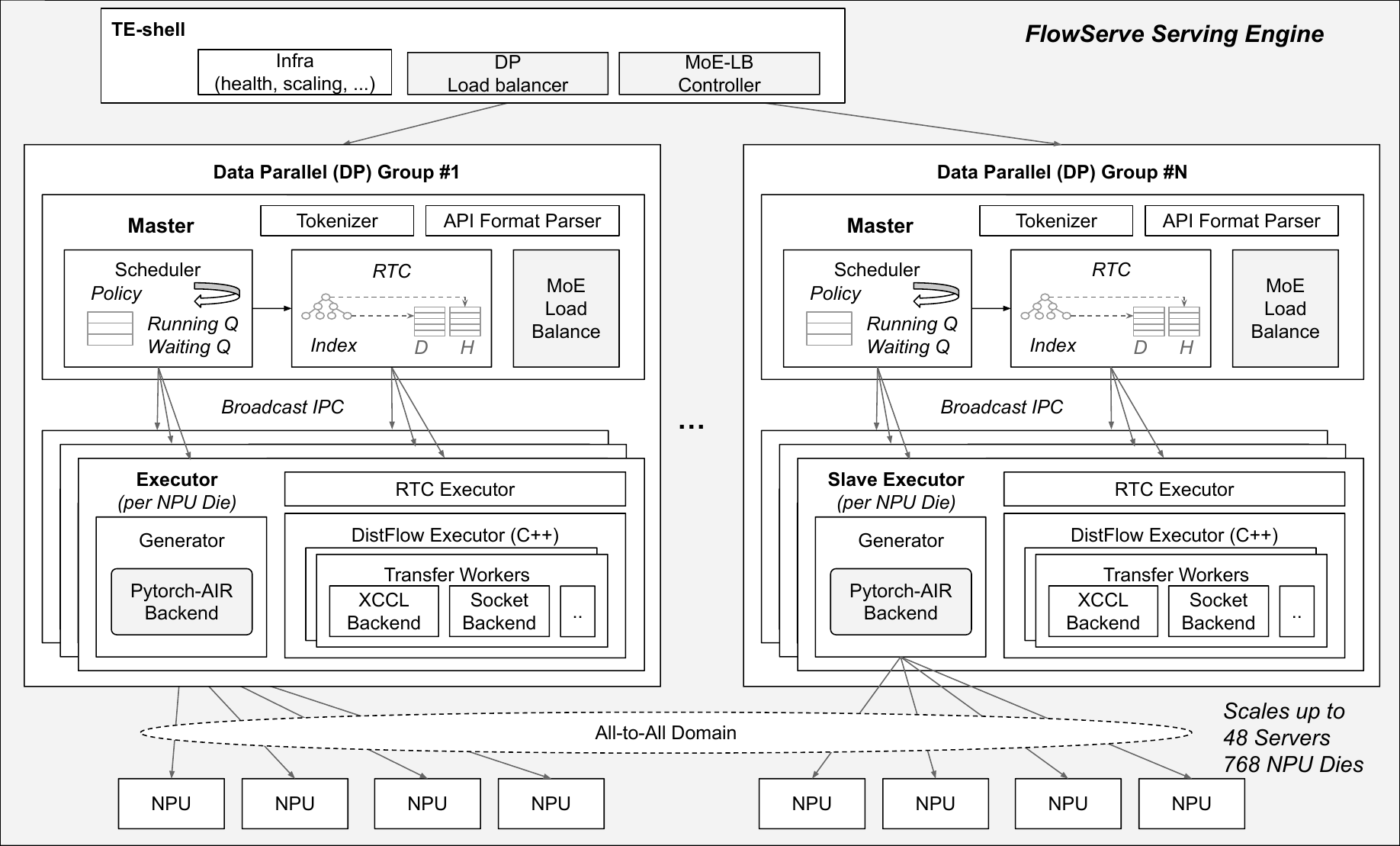}
    \caption{\textbf{The Architecture of \flowserve.}
    \textit{A single \flowserve\ engine can span an entire \cloudmatrix\ SuperPod---48 Ascend 910C servers with 768 NPU dies. \flowserve\ scales at the granularity of a DP group, where each group encapsulates a complete serving pipeline. To avoid bottlenecks and single points of failure, both request scheduling and response handling are fully distributed across DP groups.}}
    \label{fig-flowserve}
\end{figure}

\begin{figure}[t]
    \centering
    \includegraphics[width=\textwidth]{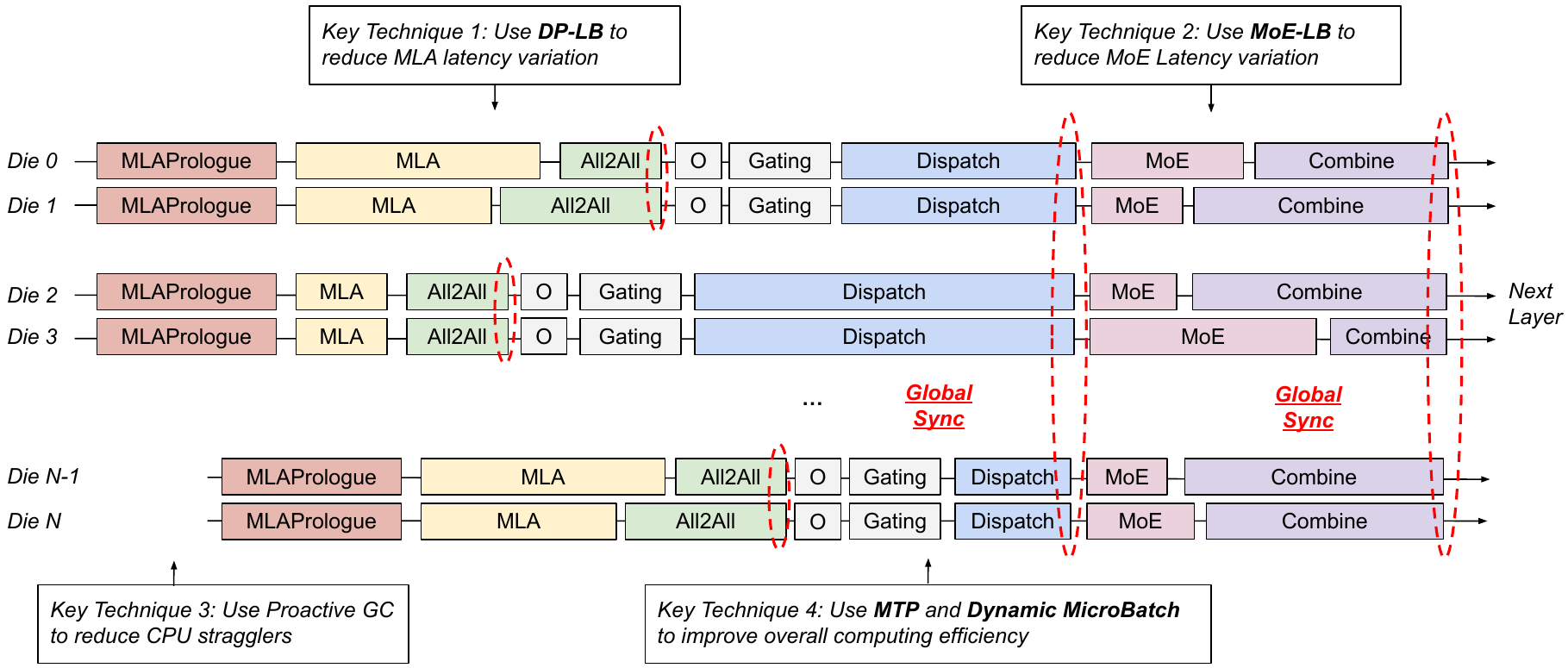}
    \caption{\textbf{The Execution Timeline of Running DeepSeek @ FlowServe.}
    \textit{This figure illustrates a single MoE layer, highlighting four key challenges it introduces along with our solutions.  
An all-to-all follows the MLA attention because we sometimes run MLAPrologue and MLA with TP=1 to avoid KV cache duplication, while executing the output projection with TP$>1$ to accelerate computation.
}}
    \label{fig-flowserve-timeline}
\end{figure}

\subsection{Overview}

The LLM model scales out via MoE; the hardware scales up via the \cloudmatrix\ SuperPod. At Huawei Cloud, our goal is to scale the serving system to efficiently run large-scale MoE models on SuperPod-scale infrastructure. Scaling introduces new challenges: it requires the right abstractions, efficient scheduling, and the elimination of both performance bottlenecks and single points of failure.

To this end, we redesign our serving
engine, \flowserve~\cite{hu2025deepflow}, evolving it from a single-node
deployment to a SuperPod-scale distributed system, as illustrated in
Figure~\ref{fig-flowserve} and Figure~\ref{fig-flowserve-timeline}. This
redesign centers on three key components:
\begin{itemize}

\item 
\textbf{First}, we introduce the \textit{Data Parallel (DP) group} abstraction, inspired by SGLang~\cite{sglang}.  
Each DP group encapsulates a full serving pipeline---including tokenization, API
parsing, SPMD-style executors, the Relational Tensor Cache (RTC), and the
DistFlow networking stack~\cite{hu2025deepflow}.  
To eliminate bottlenecks and single points of failure, both request scheduling
and response handling are fully decentralized across DP groups.

\item 
\textbf{Second}, we implement system-level optimizations to minimize latency from global synchronization.  
Although requests are scheduled independently across DP groups spanning hundreds
of NPUs, MoE models introduce two global barriers: \dispatch\ and \combine,
which can severely limit scalability.  
To address this, we apply DP-level load balancing and proactive garbage collection to reduce \dispatch\ latency variance, and MoE load balancing to optimize 
\combine.

\item
\textbf{Third}, we optimize model execution to improve forward-pass efficiency.  
We implement efficient Multi-Token Prediction (MTP) and Dynamic MicroBatching to
better utilize hardware and increase throughput when serving DeepSeek-R1/V3.

\end{itemize}

\noindent We describe these optimizations below.

\subsection{Scalable Data Parallel Group}
During the decode phase of MoE model serving, \flowserve\ adopts an
expert parallel (EP) pattern for MoE layers and a data parallel
(DP) pattern for attention layers. As the number of DPs can scale to hundreds,
a single \flowserve\ instance is capable of generating hundreds of thousands of
tokens per second.
The key challenge is to evolve \flowserve\ into a
decentralized architecture without any single point of scalability
bottleneck.

Figure~\ref{fig-flowserve} illustrates the evolved architecture of \flowserve.
The core design principle for scalability is to make each DP a self-contained
software stack and to eliminate cross-DP communication. Key components---including
the scheduler, output processing module, RTC engine, and EP
load-balancing (EP-LB) module (\S\ref{sec:ep-lb})---are replicated within each
DP. The TE-shell acts as a centralized orchestrator for cross-DP coordination,
but its responsibilities are limited to three essential functions:
dispatching requests across DPs (\S\ref{sec:dp-lb}), triggering
expert load balancing (\S\ref{sec:ep-lb}), and coordinating health
checks among all DPs (\S\ref{sec:health-check}).

In \flowserve, each request is dispatched via the DP load balancer, centralized
in the TE-shell, to achieve optimized load balancing using a global view. This
dispatch occurs only once per request, making it affordable for the TE-shell.
However, for streaming outputs, decentralization is essential. To address this,
\flowserve\ introduces an \textit{output shortcutting} optimization: the master
process of each DP spawns a separate child process dedicated to output
handling---including detokenization and output stream parsing (e.g., extracting
reasoning content or tool-call results)---and relays the resultant messages
directly to the \xds\ frontend.

\subsection{DP Load Balancing}
\label{sec:dp-lb}

In large-scale model inference systems with multiple DPs, user requests are
dispatched across several attention data parallel groups that share a common MoE
backbone. The dispatch step introduces an implicit synchronization barrier: each
DP group must wait at every MoE dispatch point for the slowest group to complete
its previous attention block computations. If attention computation times are
not evenly distributed, faster DP groups will idle, waiting for stragglers,
which increases overall tail latency at the MoE dispatch point. Therefore,
carefully balancing request routing across DP groups is crucial to prevent these
MoE synchronization delays and to maintain high throughput.

\textbf{Prefill DP Load Balancing.}
For the prefill phase, we adopt a novel single-level scheduling strategy. Our initial design extended the legacy single-DP engine to a two-level scheduler, where each request was first routed to a DP queue, and each DP ran its own local scheduler. However, this often led to stragglers---for example, one DP might pick a short batch while another handles a long one---resulting in poor utilization.
Chunk-prefill mitigates this but adds chunking overhead and requires adaptive sizing. Sequence-parallel (SP) execution posed another challenge: long requests triggered full-DP participation, but reusing the same SP ranks for both 32K and 128K sequences proved inefficient.
To balance prefill load, \flowserve\ adopts a single-level, collaborative scheduler. All tokenized requests are shared across DPs, and a leader scheduler (at \texttt{DP-0}) collects DP status via \texttt{all-gather} at each step. It then assigns request batches using a cost model (e.g., prefix cache hit rate). This global view ensures coordinated decisions across DPs.
Unlike decoupled TE-shell schedulers, this approach provides timely and accurate scheduling, invoked only when pending requests exist.

\textbf{Decode DP Load Balancing.}
In the decode phase, each DP group’s load is shaped by two factors: the number of concurrent requests it handles and the memory consumed by the KV cache.
Each DP group supports a fixed batch size. When full, it must block incoming requests until others finish, which increases both time-to-second-token (TTST) and TPOT.
At the same time, KV cache usage reflects the memory footprint of active sequences.
Uneven usage across groups leads to unbalanced MLA execution time as shown in Figure~\ref{fig-flowserve-timeline} or even triggers swapping, severely degrading performance.
To balance decode load, we first exclude DP groups that have reached their batch limit. Among the rest, we select the group with the lowest KV cache usage, accounting for reserved space needed for long outputs. The TE-shell tracks real-time metrics for each group: it updates the pending request count on dispatch and completion, and collects periodic KV cache stats. This design enables informed, system-wide scheduling decisions without introducing significant overhead.

\subsection{Proactive Garbage Collection to Reduce Jitter}
At SuperPod scale, we observe significant graph launch jitter as \flowserve's
deployment scale grows. This jitter is most pronounced at the first \texttt{dispatch}
operator (e.g., the fourth layer in DeepSeek models or the second in Kimi K2),
where global synchronization across all dies occurs for the first time. In some
cases, this jitter can exceed 100\,ms. The issue is further aggravated by
large-scale expert parallelism, where a single forward pass may involve
hundreds of NPU and CPU cores---making the system highly sensitive to stragglers.

To mitigate this jitter, we employ three key optimizations:
\begin{itemize}
  \item \textbf{Core pinning:} Each executor is pinned to a dedicated CPU core,
  minimizing kernel scheduling noise and context-switch overhead.
  \item \textbf{PTA caching:} PyTorch Air (PTA) is configured to cache compiled
  graphs, bypassing expensive runtime guard checks and reducing launch latency.
  \item \textbf{Manual Python GC:} Python’s garbage collector is invoked
  manually at controlled intervals (e.g., after every few hundred forward
  passes) to prevent unpredictable pauses during critical dispatch operations.
\end{itemize}

\begin{figure}[t]
  \centering
  \begin{subfigure}[t]{0.48\linewidth}
    \centering
    \includegraphics[width=\linewidth]{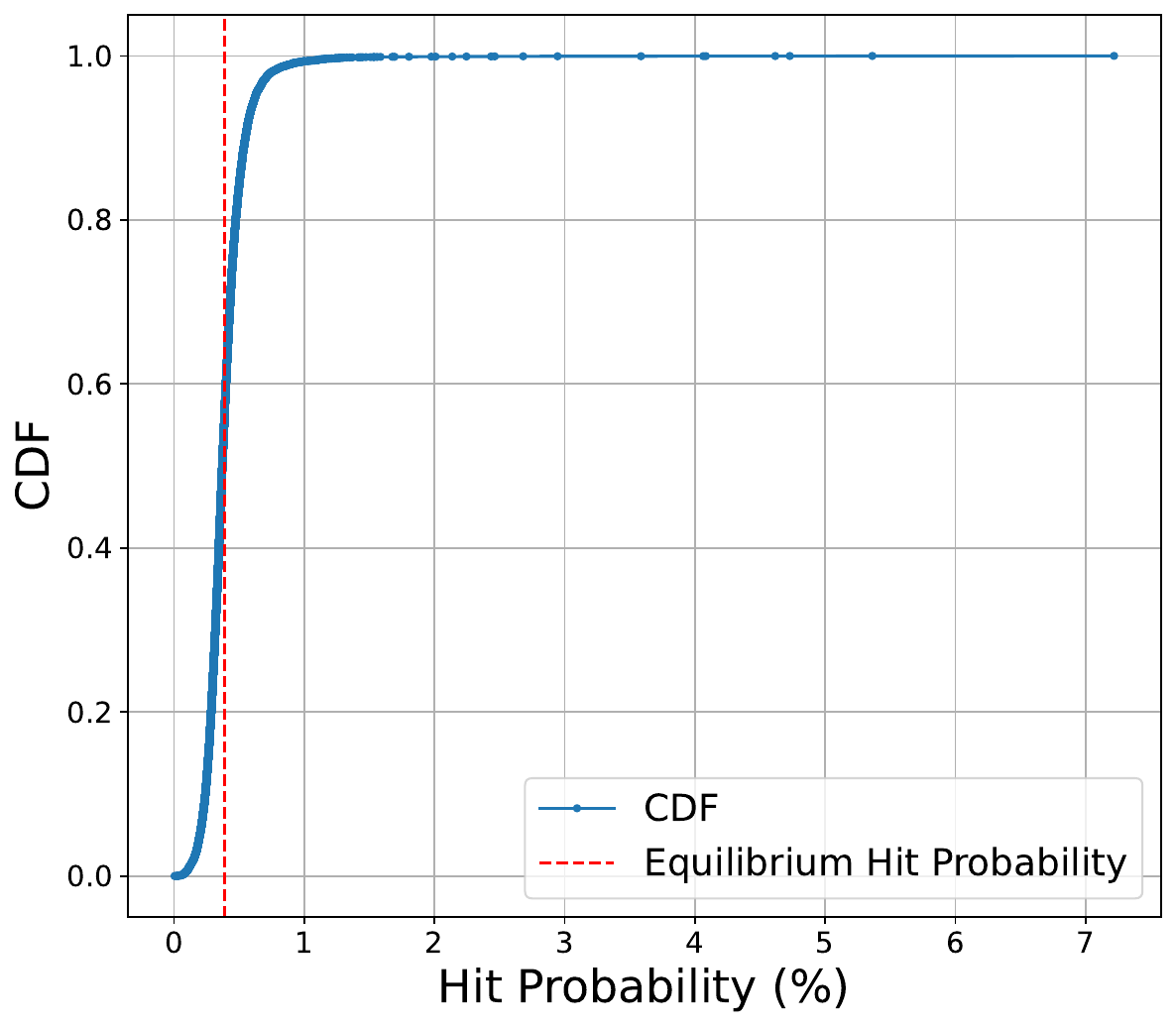}
    \caption{\textit{We show the expert load distribution of a DeepSeek-R1 layer under the ShareGPT workload. The distribution is highly skewed---20\% of experts receive more than the average load, and the hottest expert sees 30$\times$ more tokens than the average.}}
    \label{fig-moe-lb-motivation-load}
  \end{subfigure}
  \hfill
  \begin{subfigure}[t]{0.48\linewidth}
    \centering
    \includegraphics[width=\linewidth]{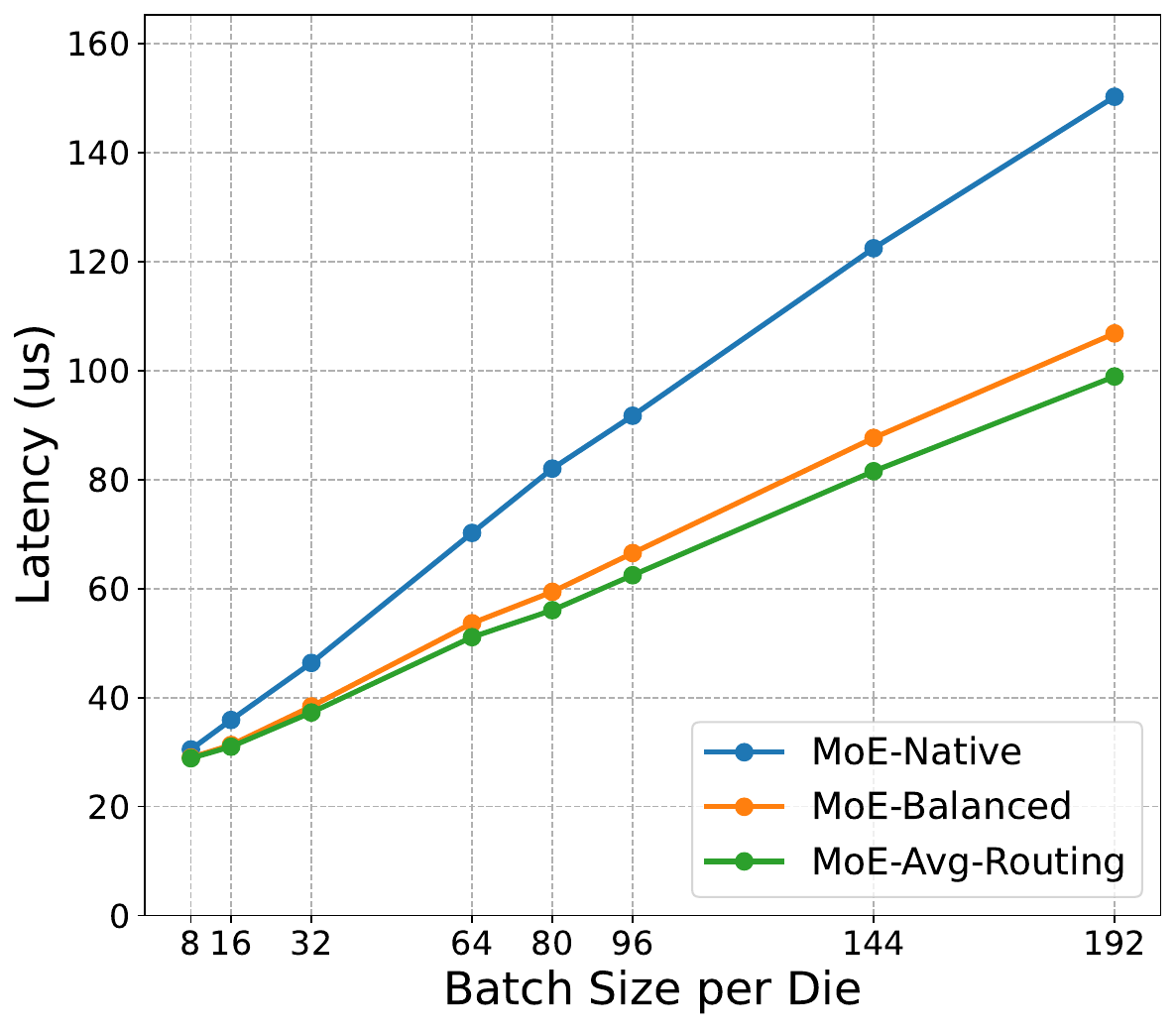}
    \caption{\textit{The setup uses EP288 and 1K-token sequence length.
    MoE-Avg-Routing, which forces uniform load across all experts; MoE-Native, which uses the original token-to-expert assignment; and MoE-Balanced, which applies our EPLB to balance expert load.}}
    \label{fig-moe-lb-motivation-perf}
  \end{subfigure}
  \caption{\textbf{A Study of Expert Placement Load Balancing}.}
  \label{fig-moe-lb-motivation}
\end{figure}

%
%
\begin{figure}[t]
    \centering
    \includegraphics[width=0.85\textwidth]{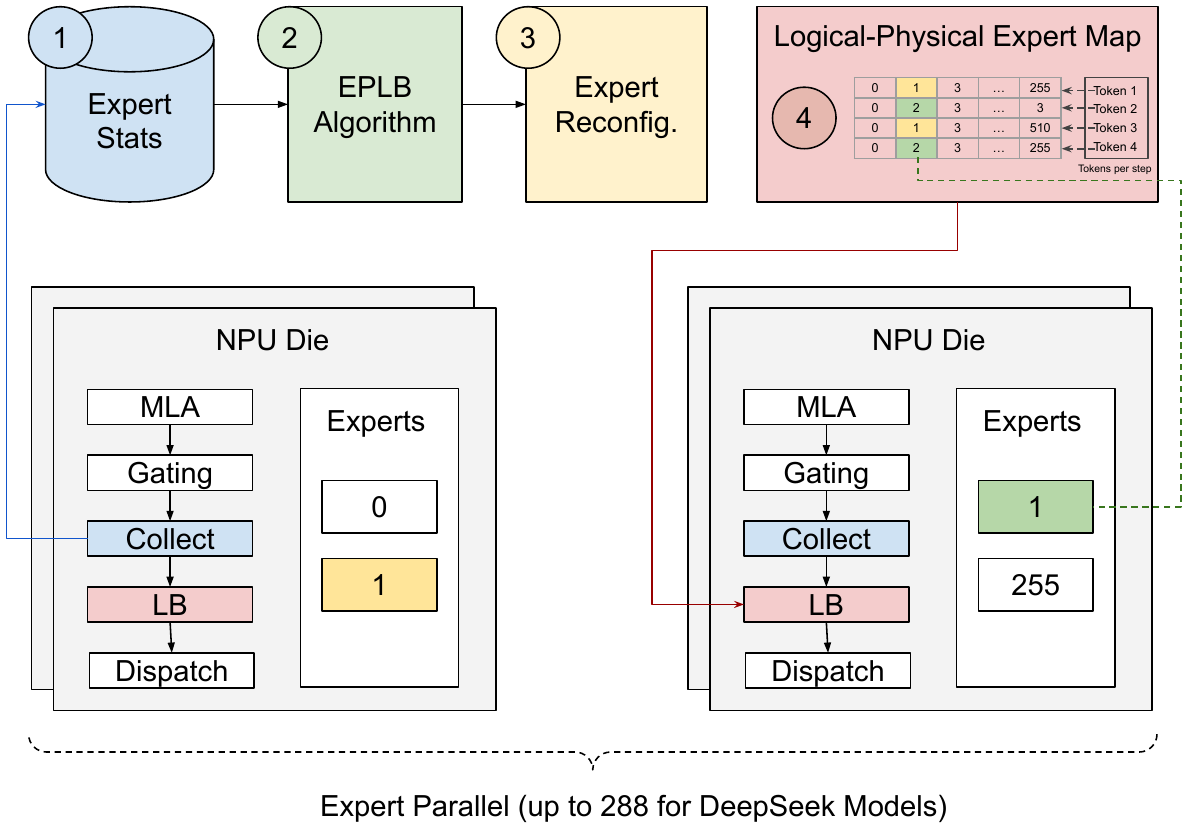}
    \caption{\textbf{An Overview of \flowserve's Expert Load Balancing.} \textit{It consists of four main components. First, expert load collection gathers token statistics across NPUs. Second, the EPLB algorithm selects redundant experts and determines their placement. Third, the system updates the expert-to-NPU mapping to reflect new replicas. Finally, during the forward pass, tokens are evenly distributed across expert replicas based on the updated mapping.}}
    \label{fig-eplb}
\end{figure}
\subsection{Expert Load Balancing}\label{sec:ep-lb}

Expert load balancing is critical for large-scale Expert Parallelism (EP) serving on \cloudmatrix. For DeepSeek models, we deploy EP288, consisting of 256 routed experts and 32 shared experts. Load imbalance among experts leads to performance degradation, causing slowdowns across all 288 NPUs due to straggler effects.
Figure~\ref{fig-moe-lb-motivation} illustrates this issue.

To address expert load imbalance, we use a data-driven approach.
Our design periodically analyzes token routing patterns, identifies overloaded (``hot'') experts, and replicates these experts across multiple NPUs. Each MoE layer reserves redundant slots per NPU to host replicated experts. During inference, we evenly distribute tokens to these replicas using precomputed mappings. Meanwhile, we continuously collect and analyze activation data to update expert replica assignments. Expert weights are swapped asynchronously to maintain uninterrupted inference throughput under significant load imbalance.

Our design includes five key components, as shown in Figure~\ref{fig-eplb}.

\noindent\textbf{Step 1: Collecting Expert Load Distribution.}
First, we collect data on expert loads across NPUs.
We define expert load as the total number of tokens routed to each expert within a given time interval.
Token count directly reflects both communication overhead (MoE-Dispatch and MoE-Combine) and computation workload (Expert MatMul).
In each MoE layer, we insert a special \texttt{Collect} kernel after gating to track the number of tokens assigned to each expert per NPU.
These counts are copied to on-chip memory's buffer.
The executor of each NPU gathers these counts within its DP group and sends the aggregated data periodically (e.g., every minute) to \flowserve's TE shell.
Frequent data collection incurs overhead, so we limit its frequency to balance accuracy and efficiency.
Once we have gathered this expert load data, we proceed to determine which experts need replication.

\noindent\textbf{Step 2: EPLB Algorithm \& Assignment.}
Given the expert load data, we use the Expert Placement Load Balancing (EPLB) algorithm to select redundant experts per layer to balance workloads.
We define the hottest expert in each time slice \(t\) for layer \(\ell\) as:
\[
h_{\ell,t} = \arg\max_{e} \text{ token\_count}[\ell][e][t].
\]

Then, the total load for layer \(\ell\) is:
\[
L_\ell = \sum_{t\in T}\text{token\_count}[\ell][h_{\ell,t}][t].
\]

Given a redundancy budget \(R\), we select redundant experts as follows:
\begin{enumerate}
    \item Compute the current total load \(L_\ell\).
    \item For each candidate expert \(c\) identified as hot in any time slice, simulate splitting tokens evenly across its replicas and compute the resulting total load \(L_\ell(c)\).
    \item Select the candidate expert \(c^*\) that minimizes the simulated load, and add it to the redundancy list.
    \item Update token counts for the selected expert to reflect even distribution among its replicas.
\end{enumerate}

With redundant experts selected, we next assign them efficiently to NPUs. To determine expert placement, we first calculate each redundant expert's total load by summing token counts across all time slices. We then sort these experts by load, starting from the highest. We assign each expert in order to the least-loaded NPU with available redundancy slots, updating that NPU's load after each assignment.

\noindent\textbf{Step 3: Redundant Expert Reconfig.}
After assigning redundant experts to NPUs, we dynamically update their configurations in four phases. First, we prefetch new expert weights from storage into memory. Second, we temporarily disable redundant expert slots by modifying the logical-to-physical expert mapping. Third, we asynchronously load the prefetched weights into the target redundant slots. Finally, we restore the logical-to-physical mapping, re-enabling the redundant expert slots. This approach ensures seamless weight updates without interrupting ongoing inference tasks.
Once reconfiguration is complete, subsequent forward passes use the updated mapping to evenly distribute tokens across expert replicas.

\noindent\textbf{Step 4: Balancing Token Loads Across Expert Replicas}  
In the final step, we ensure that tokens are evenly distributed across expert replicas during model forward. This requires mapping logical expert IDs---selected by the gating mechanism---to physical expert IDs representing actual NPU locations, while keeping both latency and communication overhead low.
Directly collecting expert activation counts from all NPUs is too expensive. To address this, we adopt two practical solutions:
\textbf{(1) Efficient Mapping with \texttt{gather} Operation.}  
We use PyTorch’s \texttt{gather} operator to efficiently map logical experts to physical replicas. This operator translates the gating matrix into actual expert assignments in parallel, minimizing computational overhead.
\textbf{(2) Communication-Free Load Balancing.}  
We implement decentralized load balancing by rotating token assignments across replicas based on each token’s position in the batch. This ensures even distribution without requiring inter-NPU communication.
As Figure~\ref{fig-moe-lb-motivation-perf} shows, we improve forward latency more than 40\%.

Figure~\ref{fig-eplb} illustrates how logical experts are mapped to rotated physical replicas based on token positions. Consider a case with 4 tokens per inference step and 256 logical experts. The mapping table has shape \textit{[4, 256]}, where each row corresponds to a token and each column to a logical expert. Suppose Logical Expert 1 has two replicas: a primary in slot 2 and a redundant in slot 1. The first column of the mapping table rotates between slots 1 and 2 across the 4 tokens, assigning each with equal probability to ensure balanced routing.


\subsection{Multi-Token Prediction}  

\begin{figure}[t]
    \centering
    \includegraphics[width=\textwidth]{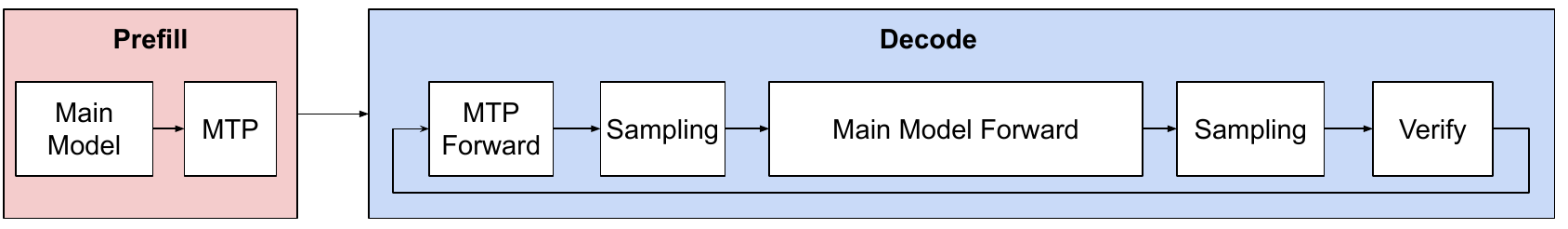}
    \caption{\textbf{An Overview of \flowserve's MTP Execution Workflow.}}
    \label{fig-flowserve-mtp}
\end{figure}

DeepSeek’s Multi-Token Prediction (MTP)~\cite{liu2024deepseek-v3} extends conventional next-token prediction by training the model to generate multiple future tokens in sequence. During inference, MTP enables speculative decoding by predicting several tokens in one pass and verifying them afterward, accelerating autoregressive generation.

This section describes how \flowserve\ integrates MTP into its disaggregated Prefill-Decode workflow (\S\ref{sec:transformerless}) to improve decode efficiency. Figure~\ref{fig-flowserve-mtp} illustrates the full execution pipeline.

\textbf{Execution Workflow.}
In the prefill stage, both the main model and MTP process the input prompt to construct the key-value (KV) cache and generate the first token. The KV cache and relevant hidden states are then transferred to the decode stage.
During decoding, we run a tightly optimized five-step loop designed to eliminate CPU-induced latency bubbles. Our initial implementation followed the EAGLE framework from vLLM~\cite{vllm-sosp23}, but its default scheduling introduced noticeable stalls. We replaced it with a customized pipeline that maximizes NPU utilization and minimizes idle time. The loop proceeds as follows:
(1) Run MTP forward to generate k draft tokens;
(2) Sample token candidates from the MTP outputs;
(3) Verify the draft tokens using the main model;
(4) Sample again from the main model outputs;
(5) Check final logits to decide token acceptance.

\textbf{Multiple MTPs.}
DeepSeek publicly released parameters for only a single MTP layer. In practice, this layer achieves a 70\%--90\% acceptance rate across most workloads, reducing latency by up to 40\% at fixed batch size.
To support deeper speculation, we initially reused the released weights for a second MTP layer without retraining. This naive setup yielded just 2.26 tokens per step on production datasets.
To improve performance, we trained a dedicated second MTP while freezing both the main model and the original MTP. Using \flowserve, we generated 280,000 samples from internal prompts. After training, the second MTP achieved 2.35 tokens per step---a 9\% improvement over the reused baseline.

\subsection{INT8 Quantization of DeepSeek Models}
\label{sec:int8-quant}

\begin{figure}[t]
    \centering
    \includegraphics[width=\textwidth]{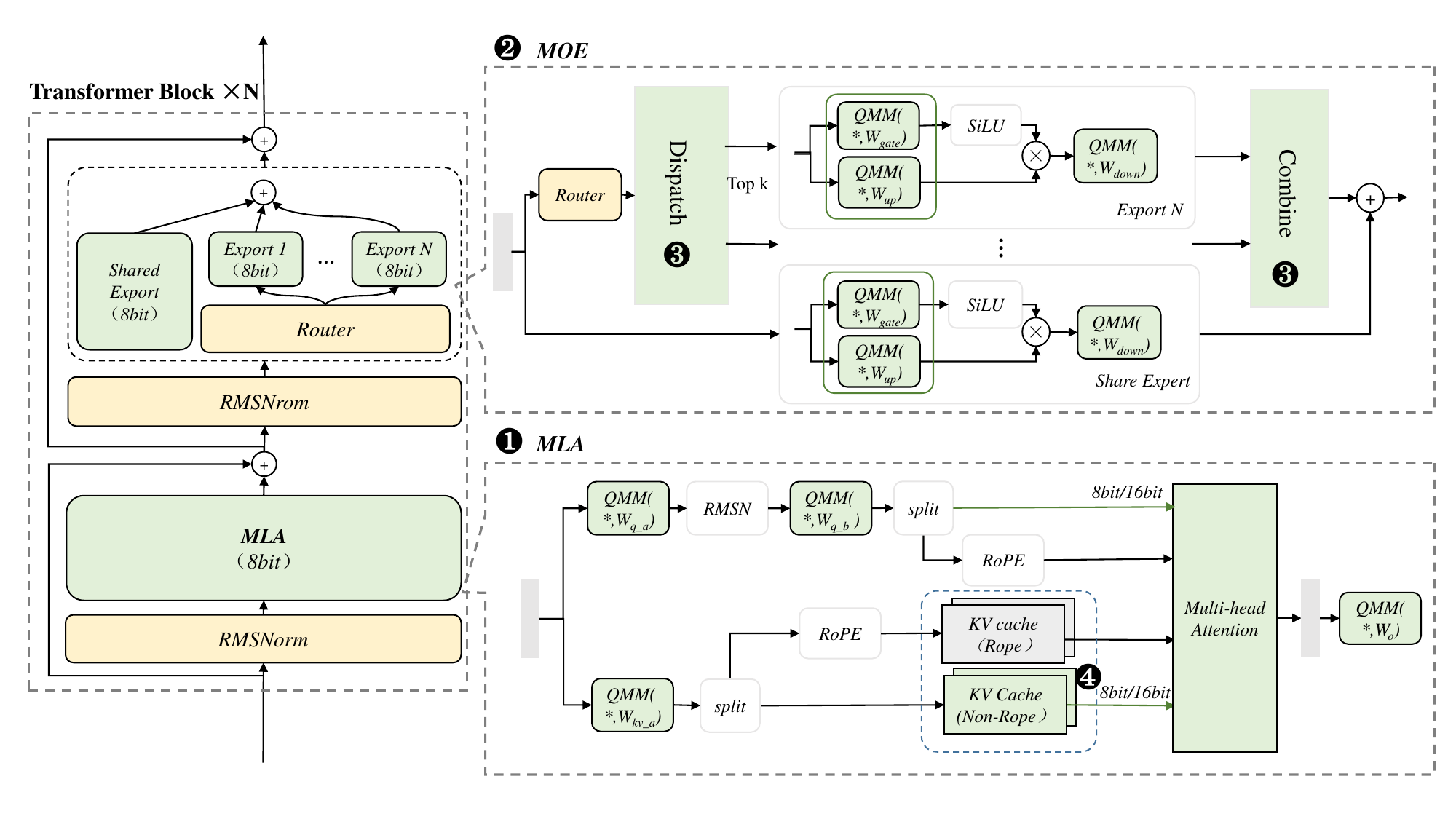}
    \caption{\textbf{An Overview of INT8 Quantization in DeepSeek Models.} \textit{We use optimized quantization strategies for MLA and MLP/MoE components, and fused quantization in MoE-dispatch communication to minimize overhead.}}
    \label{fig-quant}
\end{figure}

\begin{figure}[h]
    \centering
    \includegraphics[width=\textwidth]{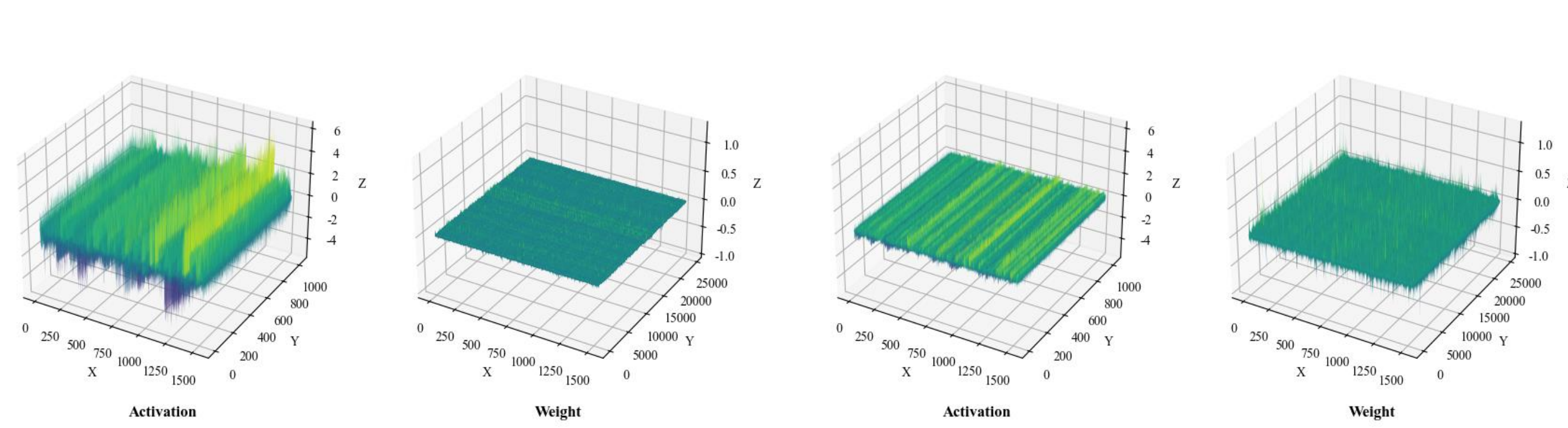}
    \caption{\textbf{Quantization Stats.} \textit{Input activation and weight magnitudes in a DeepSeek-R1 linear layer. The left two subfigures represent the distributions prior to smoothing, and the right two show how smoothing limits extreme values post-processing.}}
    \label{fig-quant-stat}
\end{figure}

The Ascend 910C NPU in \cloudmatrix\ does not natively support FP8 arithmetic. To deploy DeepSeek-R1/V3---originally trained in FP8---we quantize the model to INT8 using a Post-Training Quantization (PTQ) method.
Our method integrates SmoothQuant~\cite{smoothquant-icml23} and GPTQ~\cite{frantar2023gptq} techniques specifically for DeepSeek models.

We apply INT8 quantization to the model’s MLA, MoE, and MLP modules. To minimize accuracy loss, we introduce error compensation during quantization. Activations use token-wise quantization (one scale per token), and weights use channel-wise quantization (one scale per output channel). At inference time, we leverage the hardware-accelerated \texttt{npu\_quant\_matmul} (QMM) operator for efficient INT8 matrix multiplication.

Figure~\ref{fig-quant} summarizes our quantization approach, described in four key components:

\textbf{MLA Quantization.}  
The MLA module uses low-rank compression for the \texttt{Q}, \texttt{K}, and \texttt{V} matrices and integrates RoPE.
We quantize important weights to INT8, including query compression (\texttt{Wq\_a}), key/value compression (\texttt{Wkv\_a}), query reconstruction (\texttt{Wq\_b}), and attention output projection (\texttt{Wo}).  
Our analysis shows activations have a significantly wider dynamic range (10--100$\times$) compared to weights.
To handle this, we apply a smoothing operation that redistributes quantization difficulty between activations and weights before quantization.
Additionally, we use a GPTQ-based channel-wise quantization method with Hessian-guided iterative refinement. This approach dynamically updates the remaining FP weights to compensate for quantization errors.
Figure~\ref{fig-quant-stat} shows the input activation and weight magnitude in a DeepSeek-R1 linear layer, before and after smoothing to reduce quantization outliers.

\textbf{MLP/MoE Quantization.}  
DeepSeek models use MLP layers in early stages and MoE layers in later ones, with MoE layers---comprising both shared and routed experts---accounting for roughly 90\% of total parameters. We quantize all MLP projection weights (\texttt{up\_proj}, \texttt{gate\_proj}, \texttt{down\_proj}) and expert weights to INT8.
We apply GPTQ to dynamically update unquantized weights and reduce quantization error.
MoE expert activation patterns vary with input data, so we scale the calibration dataset to ensure each expert sees at least \(n = 4\) samples (typically 40--128) during quantization. To further improve inference efficiency, we fuse the \texttt{up\_proj} and \texttt{gate\_proj} operations into a single hardware-accelerated kernel.

\textbf{Communication Quantization.}  
To lower communication overhead, we quantize expert inputs and outputs to INT8 before sending data between nodes. Additionally, we fuse quantization and dequantization operations directly within the communication operators during inference (see \S\ref{sec:xccl-dispatch} for more detail).

\textbf{KV Cache Quantization.}  
The KV cache in MLA contains both RoPE and non-RoPE components, used to reduce computation during decoding. However, these components significantly increase memory usage for long sequences. To reduce this memory overhead, we quantize non-RoPE components---which show stable numerical distributions---to INT8. For attention layers with lower sensitivity to precision, we further optimize by performing the attention computation entirely in INT8.

\begin{figure}[t]
    \centering
    \includegraphics[width=0.95\textwidth]{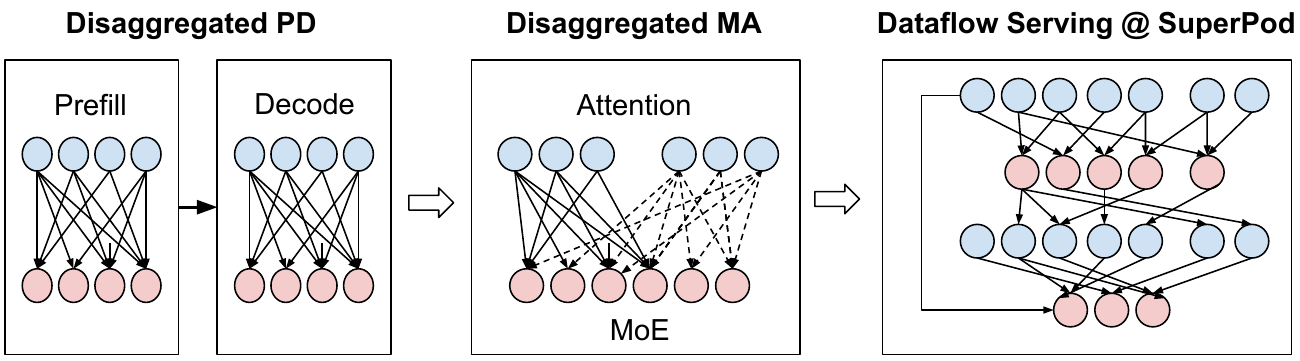}
    \caption{\textbf{The Trend Towards Transformerless Serving.} Our system evolves from a PD-colocated setting to disaggregated Prefill-Decode, then to disaggregated MoE-Attention, and ultimately into a fully asynchronous dataflow serving system.}
    \label{fig-transformerless-vision}
\end{figure}

\section{Transformerless: Towards Fully Disaggregated LLM Serving}
\label{sec:transformerless}

We advocate resource disaggregation as a core architectural principle for modern model-serving systems. Disaggregation separates system components into independently scalable units, improves fault isolation, and enables flexible evolution of system software.
Network bandwidth is the foundation of disaggregation. A 40 Gb/s scale-out network once enabled memory~\cite{farm-nsdi14} and OS disaggregation~\cite{legoos-osdi18} for traditional workloads. Today, 400 GB/s scale-up interconnects make it possible to disaggregate model serving for AI workloads.

To realize these benefits, we introduce \textbf{Transformerless}, an architecture that decomposes transformer-based LLMs into modular blocks---attention, feedforward (FFN), and MoE layers---executed independently on NPUs interconnected via high-speed fabric. The hundreds of GB/s UB interconnect in \cloudmatrix\ enables aggressive disaggregation without sacrificing inference performance.
We envision Transformerless in three stages (Figure~\ref{fig-transformerless-vision}).
First, in disaggregated Prefill-Decode (\S\ref{sec:disagg-pd}), we isolate compute-bound prefill from memory-bound decode by assigning them to different NPUs.
Second, in disaggregated MoE-Attention (\S\ref{sec:disaggg-ma}), we decouple MoE experts from attention computation, enabling independent scaling by sequence length and batch size.
Finally, in dataflow serving at SuperPod scale (\S\ref{sec:disaggg-dataflow}), we aim to break global synchronization altogether, allowing truly independent scaling of each serving component.

\subsection{Disaggregated Prefill-Decode}
\label{sec:disagg-pd}

\begin{figure}[t]
    \centering
    \includegraphics[width=0.8\textwidth]{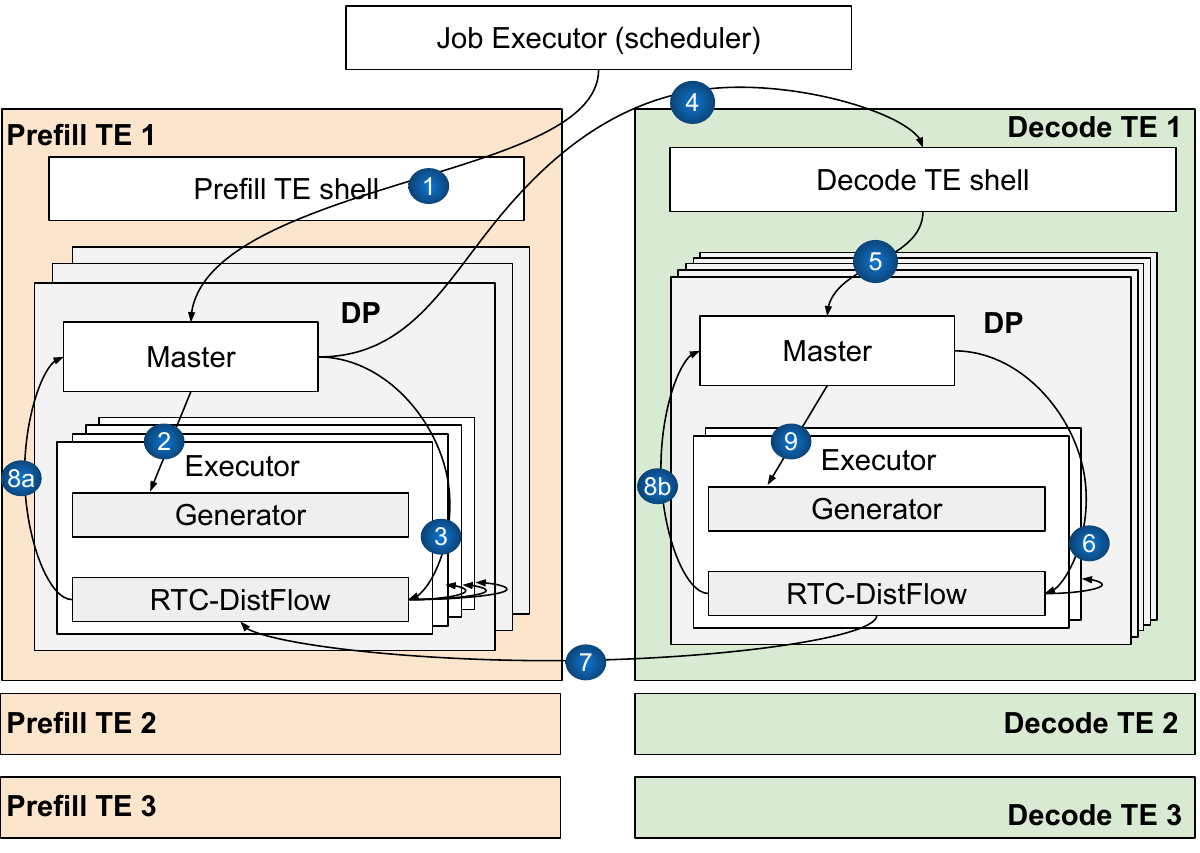}
    \caption{\textbf{The Workflow of Disaggregated Prefill-Decode over \cloudmatrix.} \textit{We support $M$ prefill and $N$ decode deployments with full-mesh connectivity. We illustrate the end-to-end workflow of sending a request from a prefill TE to a decode TE.}}
    \label{fig-flowserve-pd}
\end{figure}

Disaggregated Prefill-Decode was concurrently introduced by
Splitwise~\cite{patel2023splitwise}, TetriServe~\cite{tetriserve-arxiv24}, and
DistServe~\cite{zhong2024distserve} to reduce contention between the
compute-bound prefill and memory-bound decode phases. This design was further
refined by Mooncake~\cite{qin2024mooncake} and MemServe~\cite{hu2024memserve}.

Despite its adoption, building an efficient and robust disaggregated
Prefill-Decode pipeline remains challenging---especially for large MoE models with
expert parallelism---due to two main reasons. 
First, prefill and decode require distinct parallelization strategies because of
differing sequence lengths: we use TP=4 for prefill attention and TP=1 for
decode. This necessitates different DP groupings, making colocated execution
inefficient.
Second, the backend execution diverges: prefill uses an eager graph for
dynamic-length support, sequence parallelism, and prefix cache computation;
decode uses a static graph for maximum throughput.

We address these challenges with a fully disaggregated pipeline, as shown in
Figure~\ref{fig-flowserve-pd}.

\begin{enumerate}
    \item A request first arrives at a randomly selected Job Executor (JE),
    which assigns it to a prefill Task Executor (TE) based on a combination of
    cache status, system load, and request length. Length-awareness is crucial:
    co-locating long and short requests in the same DP group leads to
    stragglers, significantly degrading tail latency and TTFT (\S\ref{sec:dp-lb}).

    \item The prefill TE then schedules the request onto a DP group for
    computation.

    \item Upon completion of prefill, the DP master registers a PD-transfer task
	    with RTC-\allowbreak DistFlow \cite{hu2025deepflow}. This task contains only metadata
    and the addresses of the relevant KV cache blocks; actual transfer is
    deferred until triggered by the decode phase.

    \item Meanwhile, the JE dispatches the request to a decode TE, chosen based
    on real-time system load to balance throughput.

    \item The decode TE schedules the request to an appropriate DP group via
    load-aware routing.

    \item On the decode side, the DP checks for available KV block slots. If
	    capacity is insufficient, the RECV is deferred, applying backpressure to
	    upstream components. If capacity is sufficient, an asynchronous RECV is
	    submitted to DistFlow. We use the point-to-point \texttt{send}/\texttt{receive} APIs described in
	    \S\ref{sec:xccl-p2p}.

    \item DistFlow orchestrates the actual KV data transfer, handling all
    low-level concerns such as SEND/RECV handshakes, ordering guarantees, and TP
	    rank synchronization. Since KV blocks are not self-describing, DistFlow
	    ensures correct semantic pairing between prefill and decode DPs. Each
	    prefill-decode TE pair operates with an isolated DistFlow instance to limit
	    failure domains. However, multiple DistFlow instances can share \xccl\
	    buffers to optimize NPU's on-chip memory usage.

    \item Each DP polls its DistFlow completion queue. Once
    transfer completes, the prefill DP releases the KV blocks, while the decode
    DP enqueues the request for computation.
\end{enumerate}

\noindent
\textbf{Heterogeneous Prefill-Decode Deployment.}
To maximize cost-efficiency, we run prefill and decode on heterogeneous NPUs.
Prefill runs on both scale-out Ascend 910B and scale-up Ascend 910C, since it is
compute-bound and largely indifferent to interconnect bandwidth.
We place decode exclusively on Ascend 910C \cloudmatrix, where high-speed interconnect is key to meeting the 35\,ms TPOT SLA by accelerating MoE dispatch and combine.
When prefill runs on 910B and decode on 910C, we transfer KV cache over RoCE or
VPC interconnects, as shown in Figure~\ref{fig-cloudmatrix}. \flowserve\ selects
the appropriate DistFlow~\cite{hu2025deepflow} backend based on the network
fabric. For MLA models like DeepSeek and Kimi K2, both interconnects satisfy
TTFT and TPOT SLAs.

\subsection{Disaggregated MoE-Attention}
\label{sec:disaggg-ma}

\begin{figure}[t]
    \centering
    \includegraphics[width=\textwidth]{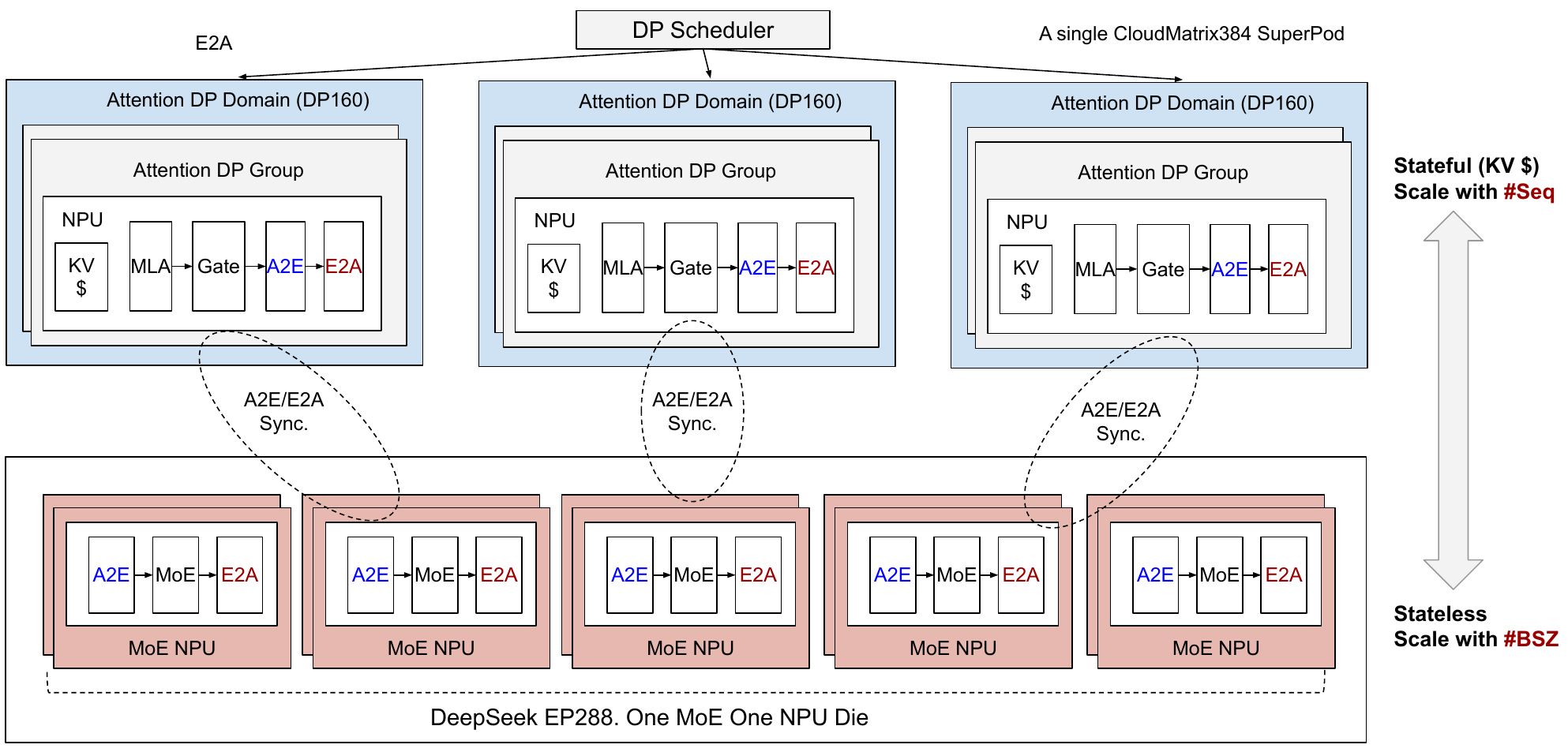}
    \caption{\textbf{The Architecture of Disaggregated MoE-Attention over \cloudmatrix.} \textit{Our deployment spans a full SuperPod with 768 NPU dies: 288 run EP288 (256 routed experts and 32 shared experts), and 480 handle MLA computation. We deploy three DP domains, each with 160 DP groups and TP=1. Attention DP groups execute full MLA computation, including MLAProlog, Attention, gating, output projection, and A2E/E2A kernels. MoE NPUs run only A2E, MoE computation, and E2A.}}
    \label{fig-flowserve-ma}
\end{figure}

\begin{figure}[t]
    \centering
    \includegraphics[width=\textwidth]{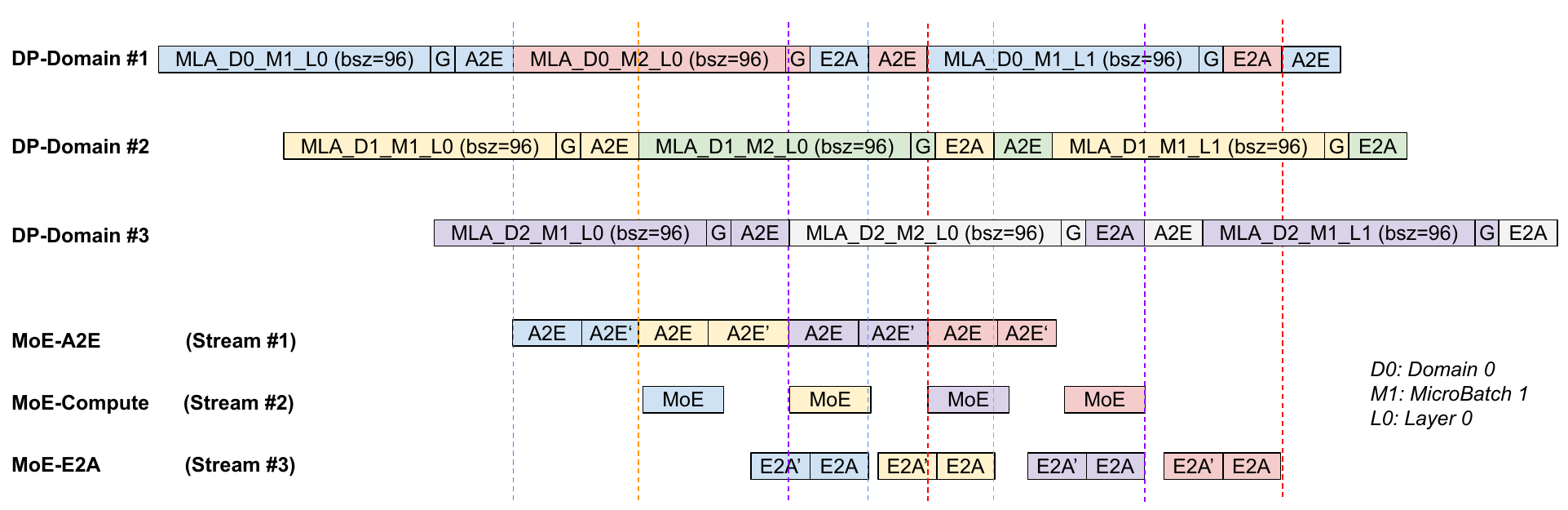}
    \caption{\textbf{The Execution Pipeline of Disaggregated MoE-Attention.} \textit{Note that each DP domain has 160 DP groups with TP set to 1. \xccl\ A2E and E2A use two-stage routing. All 288 MoE NPUs run the same persistent kernels with three concurrent streams.}}
    \label{fig-flowserve-ma-pipeline}
\end{figure}

The idea of disaggregating attention from feedforward networks (FFN) and executing them on separate resources was pioneered by works such as FastDecode~\cite{he2024fastdecode}, Lamina~\cite{chen2024efficient-ma-mingxing}, and InstAttention~\cite{pan2025instattention}, and later extended to MoE models by MegaScale-Infer~\cite{megascale-infer}.
However, prior work primarily explores small models at limited scale, leaving it unclear whether the same principles hold for large-scale MoE models like DeepSeek.

To bridge this gap, we build a large-scale disaggregated MoE-Attention system for MoE models over \cloudmatrix.  
Specifically, we run DeepSeek-V3/R1 with MoE-Attention disaggregated at the decode stage.
Our deployment spans a full \cloudmatrix\ with 768 NPU dies: 288 dies run EP288 (256 routed experts and 32 shared experts), and 480 dies run MLA.
We use DeepSeek as an example, the same design works for Kimi K2~\cite{kimi-k2} and Qwen~\cite{yang2025qwen3technicalreport}.

A key challenge for disaggregated MoE-Attention is maximizing the utilization of MoE NPUs. The MoE component is stateless, and its compute workload scales primarily with batch size, making its execution time predictable. By contrast, MLA computation (such as in DeepSeek) maintains state through KV caches and scales with both batch size and sequence length. This mismatch makes traditional approaches like microbatching or compute-communication overlap insufficient for keeping MoE NPUs fully occupied.

To solve these challenges, we introduce three techniques illustrated in Figures~\ref{fig-flowserve-ma} and~\ref{fig-flowserve-ma-pipeline}.

First, we propose new all-to-all communication primitives, \texttt{A2E} and \texttt{E2A}, tailored specifically for separate MoE and attention deployments (\S\ref{sec:xccl-a2e}). A major difficulty at large scale---such as DeepSeek-R1/V3---is asymmetric NPU allocation (e.g., 288 MoE NPUs versus 160 attention NPUs), which renders traditional pull-based mechanisms inefficient. We address this by designing a trampoline forward mechanism, where a subset of MoE NPUs (matching attention NPUs) initially receives data and subsequently forwards it to the remaining MoE NPUs. This two-stage routing reduces metadata overhead and balances bandwidth usage across NPUs, enabling efficient communication at scale. In Figure~\ref{fig-flowserve-ma-pipeline}, \texttt{A2E} and \texttt{E2A} represent the first stage, and \texttt{A2E'} and \texttt{E2A'} denote the second.

Second, we introduce a new abstraction called \textbf{Data Parallel (DP) domain}, which encapsulates multiple DP groups. A system may contain multiple DP domains, but only one domain interacts with MoE NPUs at any given time through \texttt{A2E} and \texttt{E2A} operations. Without DP domains, all DP groups would communicate concurrently, making microbatching the sole mechanism for overlapping computation and communication. However, excessive microbatching reduces the effective batch size, degrading MoE efficiency.
DP domains complement microbatching by enabling parallelism \textit{across} domains (\textbf{inter-DP parallelism}), while microbatching enables parallelism \textit{within} a domain (\textbf{intra-DP parallelism}). Figure~\ref{fig-flowserve-ma} illustrates a deployment with three DP domains, each comprising 160 DP groups (TP=1). Figure~\ref{fig-flowserve-ma-pipeline} shows the corresponding execution timeline, highlighting how DP domains and microbatching (two microbatches per domain, each of size 96) work together to improve throughput and resource utilization.

Third, we propose a \textbf{zero-overhead scheduling using persistent kernels} for MoE NPUs.
We use three concurrent streams: one for receiving data via \texttt{A2E}, one for MoE computation, and one for sending data via \texttt{E2A}. Each stream runs a persistent kernel in a busy-polling loop without returning to the CPU. This approach is critical because MoE kernels typically execute at microsecond-level granularity, and any CPU interaction (milliseconds) would introduce scheduling delays and degrade overall performance.

\subsection{Vision: Dataflow Serving at SuperPod-scale}
\label{sec:disaggg-dataflow}

No system can truly scale if global synchronizations remain in the data path.
While the disaggregated MoE-Attention architecture separates MoE and attention across NPUs, it still relies on two tightly synchronized communication primitives---\texttt{A2E} and \texttt{E2A}. As a result, a single component failure or straggler can stall the entire system, leaving it vulnerable to cascading delays.
We envision the next generation of LLM serving systems to be free of global synchronization, resembling classical dataflow architectures where tensors flow asynchronously between components. Realizing this vision poses several challenges:
First, redesigning communication protocols to tolerate latency variation without synchronous barriers.
Second, rethinking scheduling and admission control to operate in a decentralized, event-driven manner.
Third, ensuring consistency and correctness in the presence of partial results or delayed inputs.

\section{Reliability}
\label{sec:reliability}

Ensuring reliability is a key challenge when serving large-scale LLMs over
\cloudmatrix, particularly because group communication in MoE can amplify the
impact of a single-device failure into a system-wide outage.
This section presents our reliability design, focusing on high-level mechanisms.
We omit engineering details such as Kubernetes health probes, dependency
degradation strategies, and flow control mechanisms for brevity.

\subsection{Failure Detection}\label{sec:health-check}

Timely and accurate fault detection is particularly challenging in \xds, as each
\flowserve engine instance may spawn hundreds of processes across tens of nodes. While
crash failures are relatively straightforward to identify, detecting
\textit{stuck} processes---e.g., due to operators hanging on group
communication---is significantly harder.

To address this, we implement a multi-tiered heartbeat mechanism.
The \xds\ control plane periodically sends heartbeats to each \fs engine’s TE shell, which in turn forwards heartbeats to the master process of each DP group. These two intervals are decoupled for flexibility. Each DP master runs a single-threaded event loop and only responds to heartbeats when the loop is active. For instance, if an executor hangs before replying to the master, the master’s loop stalls and fails to respond---correctly indicating a fault.

However, some failure modes arise outside the main event loop. A common source
is the KV-transfer pipeline between the prefill and decode stages. This pipeline
operates asynchronously, decoupled from the main loop, and is thus invisible to
standard heartbeat checks. Failures in this path, such as a KV cache failing to
transfer, are often due to resource saturation rather than transport errors.
To detect these issues, we introduce a \textit{link probing} mechanism. By monitoring KV-transfer outcomes and injecting dummy payloads into the channel, we differentiate between decode-side saturation (which delays dummy data) and link-level faults (which block all transmission). This approach enables accurate root cause diagnosis for silent stalls and complements the heartbeat-based fault detection path.

\subsection{Failure Recovery}

Robust failure recovery is critical for large-scale LLM serving systems, where distributed execution spans hundreds of NPUs across multiple stages. Over time, we evolved our recovery strategies through three key stages---starting with coarse-grained full restarts and progressing toward fine-grained, component-level resilience. This section outlines the evolution of our recovery mechanisms, highlighting trade-offs in complexity, availability, and efficiency.

\textbf{Stage 1: Restart-the-World.}  
Our initial deployment adopted the simplest strategy: small clusters (e.g., 4 prefill and 1 decode instance, or 4P1D) and full engine restarts upon failure. When an NPU failure is detected, Kubernetes marks the node as tainted, making it unavailable to \xds. To ensure degraded clusters can still run decode instances---which span multiple nodes---we prioritize restarting decode before prefill.
While simple and reliable, this approach suffers from two key limitations. First, small clusters underutilize resources due to limited request sharing and pooling. Second, restarting the entire cluster delays recovery; a single prefill failure unnecessarily takes down all associated decode resources.

\textbf{Stage 2: Prefill/Decode (P/D) Separate Failover.}  
We next moved to shared clusters with multiple prefill and decode instances, improving resource utilization and enabling better workload specialization. For example, some prefill instances can handle long sequences to reduce head-of-line blocking, while decode capacity is pooled across requests. 
This shift required independent failover logic and introduced challenges in KV cache handling. A key issue is decode fragility---since decode spans many NPUs (e.g., 8 dies), a single failure may idle all others.
Early versions of \xds adopted a “\textbf{kill-P-to-preserve-D}” policy: prefill instances are terminated to free up resources for restarting decode. Later, we co-designed with EP-LB to support vertical scaling of decode: in failure cases, we reduce the number of DP groups and EP ranks, allowing decode to proceed with fewer NPUs. Each expert maintains at least one replica, while excess replicas are gracefully removed.

\textbf{Stage 3: Fine-Grained Error Handling.}  
Many transient failures don’t require engine restarts. We focus on two frequent cases: network instability and on-chip memory faults.
Network glitches (e.g., from switch flapping or BGP convergence) may temporarily break communication. To tolerate this, \fs adopts a \textit{token recomputation} strategy. When certain error codes are detected, all DP groups coordinate a rollback to the previous iteration. A dedicated thread broadcasts the rollback signal to ensure all DP groups---including those in busy-wait---are notified. The iteration is then re-executed, avoiding full restarts.
On-chip memory faults are another common issue. When triggered, we work with the CANN runtime to remap virtual memory and mask the faulty region. While this may cause some KV cache loss and individual request failures, the system remains online and continues serving unaffected requests.

\section{Evaluation}

\subsection{Decode Performance}

\begin{figure}[t]
\centering
\begin{minipage}{0.48\textwidth}
    \centering
    \includegraphics[width=\linewidth]{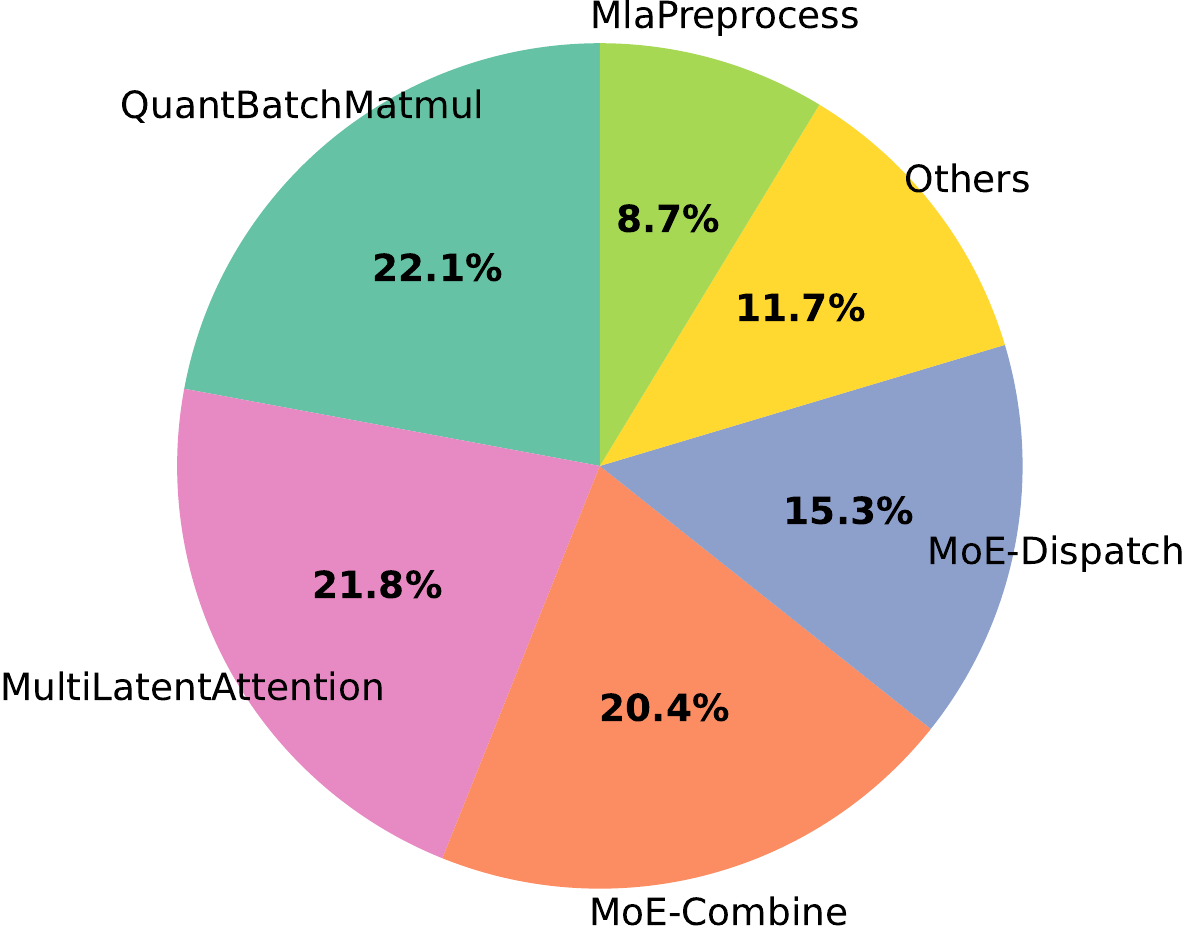} 
\end{minipage}%
\hfill
\begin{minipage}{0.48\textwidth}
	    \centering
	    {\small
	    \setlength{\tabcolsep}{4pt}
	    \begin{tabular}{lccc}
	        \toprule
	        Op      & Avg (\mus) & Min (\mus) & Max (\mus) \\
	        \midrule
	        Dispatch      & 234      & 185      & 1231     \\
	        Combine       & 312      & 165      & 2939     \\
	        \bottomrule
	    \end{tabular}
	    }
\end{minipage}
\captionof{figure}{\textbf{Latency Breakdown for One DeepSeek Decode Iteration.} \textit{Evaluation is conducted on 288 NPU dies with DP288 and EP288. Each die (or DP) uses a batch size of 60. One decode iteration---which includes MTP forward, sampling, main model forward, and final sampling---takes approximately 93\,ms. The scheduling bubble between iterations adds around 2\,ms. The average MTP acceptance rate is 90\% across all DPs. This setup achieves 2400 tokens/s per chip at a TPOT of 50\,ms.}}
\label{fig-2400-perf}
\end{figure}

We evaluate the maximum decoding performance of running DeepSeek~\cite{liu2024deepseek-v3} on \cloudmatrix.

\textbf{Setup.}  
We evaluate \flowserve\ on a single \cloudmatrix\ deployment comprising 18 Ascend 910C servers, totaling 288 NPU dies, using a \textbf{colocated prefill-and-decode setup}. The system is configured with DP288 and EP288: 256 dies host one routed expert and one redundant expert each, while the remaining 32 dies host one shared expert and one redundant expert each. We enable MTP with one MTP layer.
Each DP/die handles a local batch size of 60, yielding a global batch size of \(60 \times 288 = 17{,}280\). \flowserve\ is configured to complete prefill for all requests before entering the decode phase simultaneously. The workload uses fixed 2K-token prompts manually constructed from ShareGPT~\cite{sharegpt}, followed by 2K-token outputs. We enforce fixed-length outputs by enabling \texttt{ignore-eos}. Both the MTP module and the main model use greedy sampling during inference.

\textbf{Performance.}  
We achieve an average TPOT (time per output token) of 50\,ms. Each decode iteration---which includes MTP forward, sampling, main model execution, and final sampling (see Figure~\ref{fig-flowserve-mtp})---takes approximately 93\,ms. The scheduling gap between iterations is around 2\,ms. With an MTP acceptance rate of about 90\% across all DPs, the effective TPOT is computed as \(\frac{93 + 2}{1.9} \approx 50\)\,ms.
This translates to a per-chip throughput of \(2 \times 60 \times \frac{1000}{50} = 2400\) tokens/second (two dies per chip, each with batch size 60). For the full DP288 system, the total throughput reaches 345K tokens/second.
Figure~\ref{fig-2400-perf} shows kernel-level latency breakdown for one decode iteration (at approximately 3K sequence length). The attention kernel---scaling with both sequence and batch size---accounts for 21.8\% of latency and will grow with longer sequences. Thanks to \cloudmatrix's high-bandwidth UB fabric, \dispatch\ and \combine\ are efficient, contributing about 36\% of the total latency. However, these global synchronization kernels exhibit significant variance: their maximum latency can be up to 10$\times$ their minimum. This is because \dispatch\ absorbs variance from MLA compute across all DPs, while \combine\ absorbs imbalance from MoE compute across experts (see Figure~\ref{fig-flowserve-timeline} for execution timeline).

\textbf{Disaggregated MoE-Attention.}
We evaluate DeepSeek-R1's decoding performance using disaggregated MoE-Attention on a single \cloudmatrix\ comprising 768 NPU dies. Prefill is executed separately on other NPU servers.
The decode TE is configured as follows: 480 NPU dies are organized into three DP domains, each with 160 DP groups. The remaining 288 NPU dies run EP288.
Each DP/die processes a local batch size of 96, resulting in a global batch size of \(96 \times 3 \times 160 = 46{,}080\).
We use a fixed 2K+2K workload with \texttt{ignore-eos} enabled and greedy sampling.
Under this setup, we achieve a per-chip throughput of 2400 tokens/second with a 50\,ms TPOT.
Specifically, \texttt{MLAProlog}, \texttt{MLA}, \texttt{Gating}, and the first stage of \texttt{A2E} each take approximately 700\,ns per layer.
TPOT includes both MTP and main model forwards; the second microbatch of the final layer cannot be overlapped.
Accounting for 2\,ms scheduler overhead, 5\,ms for the MTP layer, and layer-wise compute \((0.7 \times 2 \times 61)\)\,ms, plus latencies for \texttt{A2E} (0.17\,ms), \texttt{MoE} (0.12\,ms), and \texttt{E2A} (0.19\,ms), the total forward time is approximately 93\,ms.
Given an MTP acceptance rate of 90\% across all DPs, the effective TPOT is \(\frac{93}{1.9} \approx 49\)\,ms.

\subsection{Production Workload}

We evaluate \sysname’s performance under a production workload.

A key challenge in production LLM serving is the high variability in input and output lengths, coupled with strict latency requirements. For all DeepSeek and K2 models, we target a TTFT SLA under 2\,s and a TPOT SLA of 35\,ms in most cases. In particular, for DeepSeek-R1 reasoning models, we support input sequences up to 96K tokens, with up to 32K reasoning tokens and a maximum of 32K output tokens---bounded by the model’s 128K context window.
Serving long-sequence requests differs fundamentally from short-sequence ones. Long inputs can take up to 30 minutes to process and exhibit increasing latency with sequence length, primarily due to the attention kernel’s complexity. To mitigate interference, we allocate dedicated resources to handle these extreme cases, isolating them from short-sequence traffic.

We now report the performance of a representative production setup and workload. The deployment consists of 16 Ascend 910C servers, organized into four prefill TEs and one decode TE. Each prefill TE spans two servers and is configured with DP8 and EP32. The decode TE spans eight servers with DP128 and EP128.
The workload features input lengths ranging from 0 to 64K tokens, with an average input length of 13K and an average output length of 2.1K tokens. Under this setup, we achieve a TTFT of 900\,ms and an average TPOT of 34.8\,ms.

\section{Conclusion}

\sysname\ demonstrates how to efficiently serve large-scale MoE models on SuperPod-scale infrastructure through full-system co-design.
By disaggregating transformer components, introducing a memory-semantic communication layer, and decentralizing execution with FlowServe, \sysname\ sustains high throughput and low latency across hundreds of NPUs. It runs DeepSeek models in production at 2400 tokens/s/chip while meeting a 50\,ms TPOT SLA. Looking ahead, we believe disaggregated execution will become the foundation for future LLM inference systems, especially as models and hardware continue to scale.
\section{Contributors}

Ao Xiao,
Bangzheng He,
Baoquan Zhang,
Baoxing Huai,
Bingji Wang,
Bo Wang,
Bo Xu,
Boyi Hou,
Chan Yang,
Changhong Liu,
Cheng Cui,
Chenyu Zhu,
Cong Feng,
Daohui Wang,
Dayun Lin,
Duo Zhao,
Fengshao Zou,
Fu Wang,
Gangqiang Zhang,
Gengyuan Dan,
Guanjie Chen,
Guodong Guan,
Guodong Yang,
Haifeng Li,
Haipei Zhu,
Hao Feng,
Hao Huang,
Hao Xu,
Hengrui Ma,
Hengtao Fan,
Hui Liu,
Jia Li,
Jiang Liu,
Jiang Xu,
Jie Meng,
Jinhan Xin,
Junhao Hu,
Juwei Chen,
Lan Yu,
Lanxin Miao,
Liang Liu,
Linan Jing,
Lu Zhou,
Meina Han,
Mingkun Deng,
Mingyu Deng,
Naitian Deng,
Nizhong Lin,
Peihan Zhao,
Peng Pan,
Pengfei Shen,
Ping Li,
Qi Zhang,
Qin Zhang,
Qingrong Xia,
Qingyi Zhang,
Qunchao Fu,
Ren Guo,
Ruimin Gao,
Shaochun Li,
Sheng Long,
Shentian Li,
Shining Wan,
Shuai Shen,
Shuangfu Zeng,
Shuming Jing,
Siqi Yang,
Song Zhang,
Tao Xu,
Tianlin Du,
Ting Chen,
Wanxu Wu,
Wei Jiang,
Weinan Tong,
Weiwei Chen,
Wen Peng,
Wenli Zhou,
Wenquan Yang,
Wenxin Liang,
Xiang Liu,
Xiaoli Zhou,
Xin Jin,
Xinyu Duan,
Xu Li,
Xu Zhang,
Xusheng Chen,
Yalong Shan,
Yang Gan,
Yao Lu,
Yi Deng,
Yi Zheng,
Yingfei Zheng,
Yiyun Zheng,
Yizhou Shan,
Yong Gao,
Yongqiang Yang,
Yuanjin Gong,
Yue Yu,
Yuetao Chen,
Yukun Zhu,
Yulong He,
Yusu Zhao,
Yuyan Wu,
Zenan Zhang,
Zhaojin Zhuo,
Zhaoyang Ji,
Zhefeng Wang,
Zheng Wang,
Zhenhua Yang,
Zhenli Sheng,
Zhibin Yu,
Zhigang Ji,
Zhihao Ren,
Zhipeng Bian,
Zhixia Liu,
Zhiyu Dong,
Zhonghua Li,
Zhou Yu,
Zhuoming Shen,
Zhuwei Peng,
Zi Ye,
Zihao Xiang,
Zimin Fu,
Zixuan Zhang.

\bibliographystyle{plain}
\bibliography{main}

\newpage

\end{CJK*}
\end{document}